\documentclass[a4paper,fleqn]{cas-dc}

\usepackage{amsmath,amsthm, amsfonts}
\usepackage{algorithm}
\usepackage{algpseudocode}
\usepackage{array}
\usepackage[caption=false,font=normalsize,labelfont=sf,textfont=sf]{subfig}
\usepackage{textcomp}
\usepackage{stfloats}
\usepackage{url}
\usepackage{verbatim}
\usepackage{graphicx}
\usepackage{hyperref}

\usepackage[authoryear,longnamesfirst]{natbib}

\def\tsc#1{\csdef{#1}{\textsc{\lowercase{#1}}\xspace}}
\tsc{WGM}
\tsc{QE}

\begin{document}
\let\WriteBookmarks\relax
\def\floatpagepagefraction{1}
\def\textpagefraction{.001}

\shorttitle{Thermodynamic--Complexity Duality: Embedding Computational Hardness as a Thermodynamic Coordinate}    

\shortauthors{Florian Neukart, Valerii Vinokur}  

\title [mode = title]{Thermodynamic--Complexity Duality: Embedding Computational Hardness as a Thermodynamic Coordinate}  

\tnotemark[1] 

\tnotetext[1]{} 

%

\author[1,2]{Florian Neukart}[orcid=0000-0002-2562-1618]
\author[2]{Valerii Vinokur}[orcid=0000-0002-2562-1618]

\cormark[1]

\fnmark[1]

\ead{}

\ead[url]{}

\credit{}

\affiliation[1]{organization={Gorlaeus Gebouw - BE-vleugel, Einsteinweg 55}, 
            city={Leiden},
            postcode={2333 CA}, 
            state={South Holland},
            country={Netherlands}}

\affiliation[2]{organization={Terra Quantum AG},
            addressline={Kornhausstrasse 25}, 
            city={St. Gallen},
            postcode={9000}, 
            state={St. Gallen},
            country={Switzerland}}

\cortext[1]{Florian Neukart}

\fntext[1]{f.neukart@liacs.leidenuniv.nl}


\begin{abstract}
We propose a duality between thermodynamics and computational complexity, elevating the \emph{difficulty} of a computational task to the status of a thermodynamic variable. By introducing a complexity measure \(\mathcal{C}\) as a novel coordinate, we formulate an extended first law, \(dU = T\,dS - p\,dV + \cdots + \lambda\,d\mathcal{C}\), capturing energy costs beyond classical bit erasures. This perspective unifies ideas from Landauer’s principle with the combinatorial overhead of \emph{hard} (e.g.\ NP-complete) problems, suggesting that algorithmic intractability can manifest as an additional contribution to thermodynamic potentials. We outline how this "complexity potential" might produce phase-transition-like signatures in spin glasses, random constraint satisfaction, or advanced computing hardware near minimal dissipation. We also discuss parallels with previous geometry–information dualities, emphasize the role of complexity in shaping energy landscapes, and propose experimental avenues (in reversible computing or spin-glass setups) to detect subtle thermodynamic signatures of computational hardness. This framework opens a route for systematically incorporating complexity constraints into physical modeling, offering a novel link between the fundamental cost of computation and thermodynamic laws.
\end{abstract}


\begin{highlights}
\item Developed a thermodynamic--complexity duality, treating computational difficulty as a formal thermodynamic coordinate.
\item Extended the first law of thermodynamics to include a "complexity potential," capturing energy costs of algorithmic hardness beyond basic bit erasures.
\item Demonstrated how complexity-driven corrections can manifest in phase-transition-like behavior, linking NP-hard tasks (e.g.\ SAT) to thermodynamic instabilities.
\item Established a conceptual pathway for unifying thermodynamic principles, complexity theory, and informational entropy, potentially guiding experiments in low-power/reversible computing.
\end{highlights}


\maketitle

\section{Introduction}
\label{sec:introduction}

Bringing thermodynamics and complexity theory under a unified lens not only broadens our conceptual horizons but can also reveal deep structural parallels between resource constraints in physical processes and algorithmic hardness in computation. Here, we aim to illuminate how energy, entropy, and irreversibility—central pillars of thermodynamics—can directly inform, and in turn be informed by, the formal measures of computational difficulty.

\subsection{Motivation and Scope}

Uniting thermodynamics and computational complexity into a rigorous dual framework is a challenging yet promising enterprise. In traditional physics, the laws of thermodynamics govern the flow of energy and entropy in physical systems, while complexity theory categorizes the intrinsic difficulty of computational tasks. These two fields, although developed in separate intellectual spheres, share key concepts such as \emph{entropy}, \emph{energy costs}, and \emph{limitations on process feasibility}. Over the past decades, individual connections between thermodynamics and computation have been recognized: Landauer’s principle reveals a fundamental thermodynamic price for irreversible computation \citep{Landauer61}, and Bennett’s work underscores that information processing within physical devices cannot be separated from energy dissipation \citep{Bennett82, Bennett03}. Yet, a \emph{full duality}---akin to geometrical–information dualities in certain formulations of quantum gravity \citep{VanRaamsdonk10, Swingle12, FaizalTarski2024}---remains elusive.

The analogy we aspire to construct mirrors, in spirit, the "geometry–information" frameworks that treat quantum-information measures (e.g., entanglement entropy) on equal footing with the underlying geometric degrees of freedom. Here, the proposal is to treat \emph{computational complexity} as a genuine variable capable of entering into thermodynamic potentials or laws, leading to a "complexity–thermodynamics duality." In such a duality, we might say:

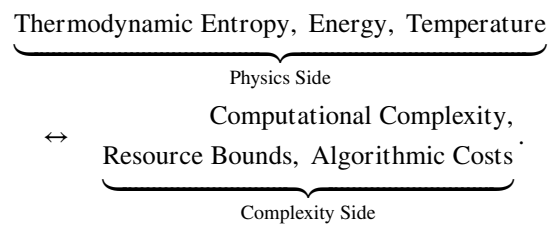
\[
\begin{aligned}
&\underbrace{\mathrm{Thermodynamic \; Entropy, \; Energy, \; Temperature}}_{\text{Physics Side}} \\
&\quad \leftrightarrow \quad
\underbrace{\begin{aligned}
\mathrm{Computational \; Complexity,} \\
\mathrm{Resource \; Bounds, \; Algorithmic \; Costs}
\end{aligned}}_{\text{Complexity Side}}.
\end{aligned}
\]

While existing studies on the thermodynamics of computation \citep{LeffRex03, Norton05} demonstrate that performing computational tasks in a physical system carries non-negligible energetic overhead, those treatments often stop short of re-deriving \emph{thermodynamic potentials} from a complexity perspective. Conversely, complexity theory rarely draws from thermodynamic potentials beyond heuristic analogies (e.g., simulated annealing \citep{KirkpatrickGelatt83}, complexity phase transitions in SAT \citep{Cheeseman91}, or free-energy analogies in spin-glass complexity \citep{MezardMontanari09}). Our goal is to push beyond heuristic parallels to formulate a unifying framework in which \emph{complexity itself modifies thermodynamic laws} (and potentially vice versa).

\subsection{What We Mean by a Duality}

In the geometry–information context (particularly in holographic correspondences), one says "information in a boundary theory \emph{is} geometry in the bulk," or that boundary entanglement patterns map onto bulk curvature \citep{RyuTakayanagi06, HubenyRangamani07}. Analogously, in a "thermodynamics–complexity duality," we want to identify:

\begin{enumerate}
    \item \emph{A Complexity Variable}---e.g., a functional $\mathcal{C}(\Phi)$ that measures the computational hardness of a certain set of tasks or states $\Phi$ within the system.
    \item \emph{A Modified Thermodynamic Potential}---e.g., a free energy $F_{\text{eff}} = U - TS + \Upsilon(\mathcal{C})$, where $\Upsilon(\mathcal{C})$ is an extra term coupling thermodynamic state variables $(U,S,\dots)$ to the complexity measure $\mathcal{C}$.
    \item \emph{Variational / Field Equations}---in which $\delta \mathcal{C}$ modifies standard thermodynamic equations or constraints on state evolutions. 
    \item \emph{Conservation or Flow Laws}---an analog of the second law that accounts for complexity as a resource or constraint.
\end{enumerate}

Importantly, we emphasize \textbf{maximal scientific depth} by grounding each step in rigorous mathematics, referencing existing knowledge in thermodynamic geometry \citep{Ruppeiner95, Weinhold75} and complexity theory \citep{GareyJohnson79, Papadimitriou94, Blum2000}.

\subsection{Historical Connections}

\paragraph{Thermodynamics of Computation.} 
Landauer’s principle, stating that erasing one bit of information dissipates at least $k_B T \ln 2$ of heat, exemplifies a "one-way link" from logic operations to thermodynamic cost \citep{Landauer61}. Subsequent work by Bennett formalized \emph{reversible computing} \citep{Bennett82}, showing how carefully arranged computational steps can asymptotically reduce energy dissipation. These approaches, however, do not typically treat \emph{complexity classes} as fundamental thermodynamic variables. They revolve around bounding or lowering the energy cost for bit operations. They address "information," but not directly "complexity" (i.e., scaling with problem size).

\paragraph{Complexity Phase Transitions and Spin Glasses.}
In spin-glass physics and random combinatorial problems, there is a well-known phenomenon of "easy–hard–easy" transitions near critical parameters \citep{Cheeseman91, MezardMontanari09}. The metaphors typically run: (1) "Energy" $\leftrightarrow$ "Cost/Objective Function," (2) "Thermodynamic Entropy" $\leftrightarrow$ "Multiplicity of Solutions," and (3) "Thermal fluctuations" $\leftrightarrow$ "Random sampling in algorithms." While powerful, these analogies remain partial. For instance, we do not see an explicit rewriting of the \emph{first law} or the partition function purely in terms of complexity classes.

\paragraph{Thermodynamic Geometry.} 
Ruppeiner geometry \citep{Ruppeiner95} and Weinhold geometry \citep{Weinhold75} introduce a \emph{metric} on thermodynamic state space. One might embed an additional "complexity coordinate," leading to an extended manifold. Variation in that coordinate could yield physically meaningful new "forces" or "fluxes," akin to your geometry–information stress-energy approach \citep{FaizalTarski2024, FreedmanHeadrick19}.

\subsection{Plan of This Work}

Our overall aim is to create a \emph{fully realized duality} between \emph{thermodynamic laws} and \emph{computational complexity}, built on the following pillars:

\begin{itemize}
    \item \textbf{Defining the Complexity Functional:} We propose a rigorous measure $\mathcal{C}(\Gamma)$, depending on the microstate or configuration $\Gamma$ of the system, capturing computational hardness or resource usage.
    \item \textbf{Constructing a Complexity-Modified Potential:} We add $\Upsilon(\mathcal{C})$ to the free energy or internal energy, then show how $\delta\Upsilon(\mathcal{C})/\delta(\dots)$ modifies standard thermodynamic relationships.
    \item \textbf{Deriving Dual Equations:} We solve new "field equations" in thermodynamic geometry that incorporate complexity, revealing a formal \emph{complexity-thermodynamics dual} in analogy with your geometry-information dualities.
    \item \textbf{Implications and Observables:} From minimal examples (like bit-erasures or NP-complete tasks embedded in a physical substrate) to extended systems (spin glasses, parallel computing structures), we demonstrate how this duality yields measurable consequences.
\end{itemize}

Moreover, we address renormalization or \emph{cutoff} issues: if $\mathcal{C}$ can become infinite for large problem sizes, an analog to entanglement regularization might be needed. We discuss how "complexity divergences" can be renormalized, shifting certain thermodynamic parameters.

\subsection{Outline of the Paper}

\noindent
\textbf{Section~\ref{sec:background_theory}} reviews necessary building blocks: classical thermodynamics, the second law, key complexity classes (P, NP, PSPACE), and previous partial connections.

\noindent
\textbf{Section~\ref{sec:duality_principle}} presents the main \emph{duality principle}, defining a "Complexity Potential" $\Upsilon(\mathcal{C})$ and showing how it modifies standard thermodynamic potentials, possibly leading to a new "Complexity Stress" in the first law.

\noindent
\textbf{Section~\ref{sec:thermo_geometry}} leverages Ruppeiner geometry to incorporate $\mathcal{C}$ as a coordinate, deriving "dual equations" reminiscent of how entanglement modifies geometry in holographic setups.

\noindent
\textbf{Section~\ref{sec:applications}} explores minimal examples, including bit erasure (Landauer limit), potential vantage on NP-complete tasks embedded in a thermodynamic system, and complexity phase transitions. We demonstrate how the dual approach yields refined predictions for energy costs or equilibrium states.

\noindent
\textbf{Section~\ref{sec:comparison}} discusses broader ramifications: open theoretical questions, potential experimental tests, synergy with quantum computing, and parallels with geometry–information dualities.

\noindent
\textbf{Section~\ref{sec:conclusion}} summarizes how a genuine "thermodynamics–complexity duality" emerges from rigorous definitions, bridging these historically separate fields into a cohesive theoretical structure.

\bigskip

\noindent
We now proceed, step by step, to develop the conceptual, mathematical, and physical scaffolding for a full-fledged thermodynamics–complexity duality.

\section{Background and Preliminaries}
\label{sec:background_theory}

This section provides the essential building blocks and notation for our subsequent derivation of a thermodynamics--complexity duality. We begin by reviewing the main elements of classical thermodynamics, focusing on the first and second laws, free-energy functionals, and the concept of thermodynamic geometry. We then move to key concepts in computational complexity theory, including complexity classes (P, NP, PSPACE) and the role of resource scaling. We conclude with a brief survey of known connections—often partial or heuristic—between these two domains.

\subsection{Thermodynamic Foundations}

\subsubsection{Zeroth, First, and Second Laws of Thermodynamics}

\paragraph{Zeroth Law.}
The zeroth law states that if two systems are each in thermal equilibrium with a third, they are in thermal equilibrium with one another. This principle underlies the definition of temperature $T$ and ensures that temperature is a well-defined transitive relation.

\paragraph{First Law.}
The first law of thermodynamics expresses the conservation of energy. In differential form,
\begin{equation}
    dU = \delta Q - \delta W \,,
\end{equation}
where $U$ is the internal energy, $\delta Q$ is the infinitesimal heat absorbed by the system, and $\delta W$ is the work done by the system. For many applications, one writes
\begin{equation}
    dU = T\,dS - p\,dV + \mu\,dN + \cdots \;,
\end{equation}
where $S$ is the entropy, $p$ is pressure, $V$ is volume, and $\mu$ is chemical potential for particle number $N$, with additional terms for other extensive variables. 

\paragraph{Second Law.}
The second law states that the entropy $S$ of an isolated system never decreases:
\begin{equation}
    dS \ge 0 \quad (\text{isolated system}),
\end{equation}
with equality holding only for reversible processes. This law underpins the \emph{arrow of time} and the irreversibility of real processes.

\subsubsection{Thermodynamic Potentials and Legendre Transforms}

In studying thermodynamic systems, one frequently employs potentials beyond the internal energy, obtained via Legendre transforms. For instance, the \emph{Helmholtz free energy} $F$ is
\begin{equation}
    F = U - T S,
\end{equation}
expressed naturally as $F=F(T,V,N,\dots)$. The \emph{Gibbs free energy} $G$, \emph{enthalpy} $H$, and \emph{grand potential} $\Omega$ are similarly defined. These potentials conveniently encode equilibrium conditions, extremized under specific constraints (constant $T$, $p$, or $\mu$).

\subsubsection{Entropy and Maxwell Relations}

From a differential perspective, each thermodynamic potential $X$ (like $F$ or $G$) has a total differential 
\begin{equation}
    dX = -S\,dT - p\,dV + \mu\,dN + \cdots,
\end{equation}
yielding the Maxwell relations by cross-partial derivatives. These relations encode the structure of thermodynamic response functions (heat capacities, compressibilities, etc.). They will be analogized later when we introduce a "complexity coordinate" that might produce new cross-derivative identities.

\subsubsection{Landauer’s Principle and Minimum Energy Cost of Information Erasure}

Although classical thermodynamics is typically formulated in purely macroscopic variables, Landauer’s principle shows that \emph{erasing one bit of information} in a system at temperature $T$ requires a minimum energy cost of
\begin{equation}
    \Delta E \geq k_B T \ln 2,
\end{equation}
where $k_B$ is Boltzmann’s constant \citep{Landauer1961}. This result embodies the partial link between logical operations (information) and thermodynamic cost, but does not by itself define a measure of \emph{computational complexity} as a variable within thermodynamics. We will later embed this principle in a broader perspective where complexity classes may impose energy constraints.

\subsubsection{Thermodynamic Geometry (Weinhold \& Ruppeiner)}

To treat thermodynamic states as points in a manifold, one introduces a \emph{metric} $g_{\alpha\beta}$ on the space of extensive or intensive variables \citep{Weinhold1975, Ruppeiner1995}. For instance, Ruppeiner’s metric is often defined via the Hessian of the entropy $S(U,V,\dots)$:
\begin{equation}
    g_{\alpha\beta}^{(\mathrm{Rup})} = -\frac{\partial^2 S}{\partial X^\alpha\,\partial X^\beta},
\end{equation}
where $\{X^\alpha\}$ are thermodynamic coordinates. Geodesics in this manifold can correspond to optimal paths of fluctuation. In a complexity–thermodynamics dual, we plan to add a new variable $\mathcal{C}$, rewriting $S(U,V,\mathcal{C},\dots)$ or $F(T,V,\mathcal{C},\dots)$, and analyzing how curvature changes.

\subsection{Computational Complexity Theory}

\subsubsection{Foundational Complexity Classes: P, NP, PSPACE}

\paragraph{P (Polynomial Time).}
Class P comprises decision problems solvable by a deterministic Turing machine in time polynomial in the size of the input, e.g., $O(n^k)$ for some $k$. Problems in P are traditionally viewed as "efficiently solvable" \citep{Garey1979}.

\paragraph{NP (Nondeterministic Polynomial Time).}
A language is in NP if, given a purported solution (certificate), one can verify correctness in polynomial time. While enumerating solutions may be exponential, verifying a guess is quick. The famous "P vs. NP" question asks whether every NP problem also lies in P.

\paragraph{PSPACE.}
This class denotes problems solvable with a polynomial amount of memory, ignoring time constraints. PSPACE subsumes NP and P, potentially introducing even more complex tasks that can still be polynomially bounded in space usage.

\subsubsection{Complexity Measures: Circuit Complexity, Kolmogorov Complexity, \emph{etc.}}

\paragraph{Circuit Complexity.}
One measure of hardness is the minimal Boolean circuit size required to decide a language. Alternatively, monotone circuit complexity can be relevant in certain spin-glass or SAT analogies \citep{Blum2000}.

\paragraph{Kolmogorov Complexity.}
Kolmogorov complexity $\mathrm{K}(\text{string})$ is the length of the shortest description of that string on a universal Turing machine. Though it addresses "descriptive complexity," it also ties to \emph{uncomputability}, so direct usage in thermodynamics might require carefully chosen approximations.

\paragraph{Resource-Bounded Measures.}
We can define time- or space-bounded complexity $\mathcal{C}(\Phi)$ for a configuration $\Phi$, capturing the minimal resource cost to solve or decide something about $\Phi$. In a thermodynamic system, $\Phi$ might represent microstates, boundary conditions, or constraints, from which we measure computational hardness.

\subsubsection{Phase Transitions in Complexity: Spin Glasses, Random SAT, \emph{etc.}}

Several well-known phenomena highlight "hardness peaks" in random combinatorial problems as system parameters (e.g., clause-to-variable ratio in SAT) cross a critical threshold \citep{Cheeseman1991, Mezard2009}. This parallels second-order phase transitions in statistical physics, albeit the precise mapping to free energy is typically heuristic. One ambition of the dual framework is to recast these complexity thresholds in a rigorous thermodynamic potential or an extended manifold.

\subsection{Known Intersections of Thermodynamics and Computation}

\paragraph{Thermodynamics of Computation.}
We have already mentioned Landauer’s principle \citep{Landauer1961} and Bennett’s reversible computing \citep{Bennett1982, Bennett2003}. They highlight a partial interplay: \emph{information} $\to$ \emph{heat}, but do not treat complexity classes per se.

\paragraph{Simulated Annealing.}
The celebrated optimization heuristic \citep{Kirkpatrick1983} draws an analogy between searching for global minima in a cost landscape and allowing a thermodynamic system to cool slowly. However, it remains a technique rather than a fundamental duality. It does not incorporate a "complexity variable" or yield exact modifications to $dU = T\,dS - p\,dV$.

\paragraph{Complexity at Phase Transitions.}
Work on "algorithmic complexity near spin-glass transitions" has found that the hardest instances align with phase boundaries \citep{Cheeseman1991}. Yet again, the partial analogy stops short of rewriting thermodynamic potentials in complexity-theoretic terms or introducing new conjugate variables.

\subsection{Summary of the State of the Art}

To date, the "thermodynamics of computation" is extensively studied regarding \emph{bit-level} heat costs and the possibility of reversible gates. Meanwhile, "complexity theory" incorporates occasional thermodynamic metaphors but seldom modifies fundamental laws of thermodynamics. A deeper unification---introducing a \emph{complexity measure} as a legitimate thermodynamic coordinate and deriving new "dual equations"---has not yet emerged.

In the next section, we propose precisely such a unifying strategy: define a "complexity potential," incorporate it into thermodynamic geometry, and show how the usual laws of thermodynamics become augmented by complexity terms in a manner that parallels entanglement–geometry dualities in quantum gravity contexts.

\section{The Duality Principle: Toward a Complexity-Modified Thermodynamics}
\label{sec:duality_principle}

Having laid out the fundamental concepts in both classical thermodynamics and computational complexity, we now formulate the \emph{core duality principle}: we treat \emph{computational complexity} as an intrinsic variable in a thermodynamic description, thereby modifying and extending the usual laws and potentials of thermodynamics. This section presents the mathematical structure underlying such an approach, starting by defining a \emph{complexity functional} and then introducing an additional term in the free energy that depends on this complexity. We also explore how variations of this new term lead to modified Euler relations and Maxwell-like identities. Our treatment here parallels the role of entanglement entropy as a geometric variable in holographic setups, but with computational hardness in place of quantum correlations.

\subsection{Defining a Complexity Functional: \texorpdfstring{$\mathcal{C}$}{C}}

\subsubsection{General Form of $\mathcal{C}$}
We posit a complexity measure $\mathcal{C}$, which depends on the system's configuration or microstate $\Gamma$. Formally,
\begin{equation}
    \mathcal{C}(\Gamma) : \; \Omega \;\to\; \mathbb{R}_{\ge 0},
\end{equation}
where $\Omega$ is the space of physically realizable states. The dimensionless real value $\mathcal{C}(\Gamma)$ is intended to capture the intrinsic difficulty of performing certain computational tasks \emph{on or with} $\Gamma$. For instance, $\Gamma$ might encode constraints or boundary conditions that define a decision problem or optimization scenario. A higher value of $\mathcal{C}$ indicates a more intractable or resource-intensive computational challenge.

\paragraph{Finite vs.\ Unbounded Complexity.}
In some contexts, $\mathcal{C}$ can grow unboundedly with system size $N$ (for example, $2^N$ for enumerative tasks). To embed such divergences in a thermodynamic framework, we may adopt:
\begin{itemize}
    \item \emph{Scaling forms:} $\mathcal{C}(\Gamma) \sim \exp(\alpha N)$ or $\sim N^k$, to reflect exponential or polynomial growth.
    \item \emph{Regularization} or \emph{renormalization} strategies (cf.\ Sec.~\ref{sec:regularization} below), wherein large-$N$ divergences in $\mathcal{C}$ are absorbed into redefinitions of certain thermodynamic parameters.
\end{itemize}

\subsubsection{Analogy with Entanglement Entropy}
In quantum gravity or holography, one treats entanglement entropy $S_{\text{EE}}$ as a variable that can backreact on geometry. Here, $\mathcal{C}$ is an \emph{algorithmic} or \emph{computational} measure, but we aim for an analogous role: $\mathcal{C}$ will act as a coordinate in an extended thermodynamic manifold.

\subsubsection{Time/Space Complexity or Circuit Depth?}
We leave $\mathcal{C}$ general: it might represent time complexity for some decision problem embedded in $\Gamma$, or circuit size, or even problem-specific complexity (like a SAT instance’s hardness). The existence of multiple complexity classes (P, NP, PSPACE, \dots) suggests different functional forms of $\mathcal{C}$. Our formalism works for a broad range of such definitions, provided $\mathcal{C}$ is suitably smooth in its dependence on thermodynamic parameters (see the discussion in Sec.~\ref{sec:regularization}).

\subsection{Introducing a Complexity Potential \texorpdfstring{$\Upsilon(\mathcal{C})$}{Upsilon(C)}}

\subsubsection{Augmented Free Energy}
To embed $\mathcal{C}$ into thermodynamics, consider the Helmholtz free energy
\begin{equation}
    F = U - T S,
\end{equation}
with natural variables $(T, V, N, \dots)$. We define an augmented free energy (or "complexity-modified free energy"):
\begin{equation}
    F_{\text{eff}}(T, V, N, \mathcal{C}) \;=\; F(T, V, N)\;+\;\Upsilon\bigl(\mathcal{C}\bigr),
\end{equation}
where $\Upsilon(\mathcal{C})$ is a new scalar function that depends on the complexity measure $\mathcal{C}$. This extra term effectively couples complexity to the thermodynamic degrees of freedom.

\paragraph{Interpretation.}
In normal thermodynamics, $F$ is minimized at equilibrium for a system at fixed $T, V, N$. Now, if $\mathcal{C}$ varies, the stationarity conditions for $F_{\text{eff}}$ may require an \emph{optimal complexity} $\mathcal{C}_{\text{eq}}$. The physical meaning is that the system might "choose" or be constrained to a certain complexity if it can reduce overall free energy cost. This is reminiscent of how quantum circuits in the AdS/CFT dual interpret "circuit depth" as a gravitational action \citep{Chapman2018}.

\subsubsection{Units and Dimensional Analysis}
We require $\Upsilon(\mathcal{C})$ to share the units of free energy (e.g., Joules in SI). Meanwhile, $\mathcal{C}$ is dimensionless or measured in bits. Thus, the simplest dimensionally consistent possibility is 
\[
    \Upsilon(\mathcal{C}) \;=\; k_B T_0 \, \varphi(\mathcal{C}),
\]
where $T_0$ is some reference temperature (or energy scale) ensuring correct energy units, and $\varphi$ is dimensionless. More refined scalings could exist, especially if we let $\Upsilon$ vary with other control parameters (like $T$ or $N$). For now, we take $T_0$ as a fixed constant.

\subsection{Modified Euler Relations and New Conjugate Variable}

\subsubsection{Differential Form of \texorpdfstring{$F_{\text{eff}}$}{F eff}}
The total differential of $F_{\text{eff}}$ is
\begin{align}
dF_{\text{eff}} &= dF + d\Upsilon(\mathcal{C}) 
= dF + \frac{d\Upsilon}{d\mathcal{C}}\,d\mathcal{C}
\nonumber \\
&= -S\,dT - p\,dV + \mu\,dN \;+\; 
\Psi \, d\mathcal{C}.
\end{align}
Here, we define
\begin{equation}
    \Psi \;=\; \frac{d\Upsilon}{d\mathcal{C}},
\end{equation}
which plays the role of a \emph{conjugate variable} to $\mathcal{C}$ in the extended thermodynamic manifold. By analogy with "$p$ is conjugate to $V$, $T$ is conjugate to $S$," we interpret $\Psi$ as "energy cost per unit complexity change," or "chemical potential for complexity."

\subsubsection{Euler Relation and Gibbs--Duhem-Like Constraints}
From the Euler or Gibbs--Duhem perspective, standard thermodynamics yields relations like
\[
F = -p\,V + \mu\,N + \ldots \quad (\text{for certain homogeneous systems}).
\]
When $\mathcal{C}$ is added, we may get an extended Euler equation of the form
\begin{equation}
F_{\text{eff}} \;=\; -p\,V + \mu\,N + \ldots + \Psi\,\mathcal{C}.
\label{eq:extended_Euler}
\end{equation}
This equality may hold if $F_{\text{eff}}$ is homogeneous of degree 1 in the extensive variables (including $\mathcal{C}$ if it is treated as "extensive" in some sense). The precise form depends on whether $\mathcal{C}$ is regarded as extensive (growing with system size) or intensive (like an average cost). 

\paragraph{Consistency with the Second Law.}
If the system can freely adjust $\mathcal{C}$, then from $dF_{\text{eff}} = 0$ at equilibrium, we get
\[
\Psi \, d\mathcal{C} + \ldots = 0,
\]
suggesting that $\mathcal{C}$ might settle in a certain value so as to minimize $F_{\text{eff}}$. Alternatively, if $\mathcal{C}$ is externally specified (like an externally imposed "hardness level"), then $\Psi$ is determined by the system’s response.

\subsection{Mathematical Parallels to Entanglement-Driven Equations}

\subsubsection{Analogy to Maxwell Relations}

From $dF_{\text{eff}}$, we can read partial derivatives:
\begin{equation}
\begin{aligned}
- \left(\frac{\partial F_{\text{eff}}}{\partial T}\right)_{V,N,\mathcal{C}} &= S, \quad
- \left(\frac{\partial F_{\text{eff}}}{\partial V}\right)_{T,N,\mathcal{C}} = p, \\
\left(\frac{\partial F_{\text{eff}}}{\partial \mathcal{C}}\right)_{T,V,N} &= \Psi.
\end{aligned}
\end{equation}
Cross-derivative consistency yields new Maxwell-like identities:
\begin{equation}
\begin{aligned}
\frac{\partial S}{\partial \mathcal{C}}\Big|_{T,V,N} 
&= \frac{\partial \Psi}{\partial T}\Big|_{V,N,\mathcal{C}},\\
\frac{\partial p}{\partial \mathcal{C}}\Big|_{T,N} 
&= \frac{\partial \Psi}{\partial V}\Big|_{T,N,\mathcal{C}},
\quad \dots
\end{aligned}
\end{equation}
These are reminiscent of how "entanglement chemical potential" leads to cross-terms in holographic entanglement thermodynamics \citep{Takayanagi2018Analogies}. Here, $\Psi$ is the "complexity potential."

\subsubsection{Effective Partition Function or Complexified Partition}

If one re-derives partition functions from a path-integral perspective (in a classical or quantum setting), the presence of $\Upsilon(\mathcal{C})$ modifies the Boltzmann factor $\exp(-\beta F) \to \exp[-\beta F - \beta \Upsilon(\mathcal{C})]$. Symbolically, one might write
\[
Z_{\text{eff}}(T,\ldots) \;=\; 
\int d\Gamma \,\exp\Bigl(-\beta H(\Gamma)\Bigr)\,\exp\Bigl(-\beta \Upsilon(\mathcal{C}(\Gamma))\Bigr).
\]
Though more formal development is needed to interpret how $\mathcal{C}(\Gamma)$ is embedded in a Hamiltonian or path integral, this approach could unify the "thermodynamic sum over microstates" with an "algorithmic cost factor." The stationary condition on $\mathcal{C}$ might lead to a saddle-point argument reminiscent of large-$N$ expansions in field theory.

\subsection{Regularization and Divergence Control}
\label{sec:regularization}

As with entanglement entropy, which often has ultraviolet divergences requiring a cutoff, complexity $\mathcal{C}$ may become arbitrarily large with system size or problem constraints. A naive attempt to incorporate such a diverging $\mathcal{C}$ might produce infinite shifts in $F_{\text{eff}}$. To handle this:

\begin{enumerate}
\item \textbf{Introduce a Complexity Scale} $\mathcal{C}_0$ analogous to a UV cutoff, ensuring that only "excess complexity" $\mathcal{C}-\mathcal{C}_0$ enters the potential.  
\item \textbf{Renormalize $\Upsilon(\mathcal{C})$} by subtracting infinite or large constant pieces, defining $\Upsilon_{\text{ren}} = \Upsilon - \Upsilon_{\infty}$.  
\item \textbf{Absorb Divergences into Coupling Constants} in the manner of quantum field theory: e.g., $\Upsilon(\mathcal{C})$ might shift certain thermodynamic parameters (like $G$ in a gravitational analogy) \citep{LevinFaizal2025Hypothetical}.
\end{enumerate}

Thus, one obtains finite physical predictions. The net effect: we treat $\mathcal{C}$ divergences similarly to how entanglement entropy divergences are renormalized.

\subsection{Physical and Interpretational Remarks}

\paragraph{Interpretation of $\Psi$.}
Just as $p = -(\partial F/\partial V)$ measures how free energy changes with volume, $\Psi = (\partial F_{\text{eff}}/\partial \mathcal{C})$ indicates how overall system free energy changes if the complexity is marginally varied. If $\mathcal{C}$ is externally imposed, $\Psi$ is like a "Lagrange multiplier" enforcing that complexity constraint.

\paragraph{Mechanical Analogy.}
One might imagine $\mathcal{C}$ as an extra coordinate in a classical "thermodynamic machine." The work done in changing complexity is $\Psi \,\mathrm{d}\mathcal{C}$, akin to $p\,\mathrm{d}V$. In an actual computing device, $\Psi \,\mathrm{d}\mathcal{C}$ might represent extra energy cost needed for more demanding tasks.

\paragraph{Parallels with AdS/CFT Complexity.}
In holography, the "complexity = action" or "complexity = volume" proposals treat certain geometric volumes or gravitational actions as proxies for circuit complexity in the dual boundary theory \citep{Chapman2018}. One obtains analogs of "thermodynamic relations" with $\mathcal{C}$ as a variable. Our approach is reminiscent of this but purely in classical thermodynamics with an explicit $\Upsilon(\mathcal{C})$ term.

\subsection{Summary of the Duality Construction}

We have introduced the augmented free energy $F_{\text{eff}} = F + \Upsilon(\mathcal{C})$. From its differential form,
\[
    dF_{\text{eff}} = -S\,dT -p\,dV + \mu\,dN + \Psi\,d\mathcal{C},
\]
we identify $\Psi$ as conjugate to $\mathcal{C}$. This yields new Maxwell-like relations and a potential extension of standard Euler relations, bridging \emph{complexity} with \emph{thermodynamic} variables. The extent to which $\mathcal{C}$ is dynamically chosen or externally controlled depends on physical modeling. The next step (in Sec.~\ref{sec:thermo_geometry}) is to show how $\mathcal{C}$ enters a thermodynamic geometry approach, generating new curvature components that mirror the role of entanglement in holographic dualities.

\section{Thermodynamic Geometry of Complexity}
\label{sec:thermo_geometry}

Thus far, we have presented a duality principle in which a complexity measure \(\mathcal{C}\) is introduced as an additional thermodynamic variable, alongside a new potential \(\Upsilon(\mathcal{C})\). Building on that foundation, we now develop a \emph{geometric} interpretation reminiscent of Weinhold and Ruppeiner formalisms. Specifically, we formulate a \emph{metric tensor} that extends thermodynamic geometry to include complexity, enabling us to investigate concepts such as curvature, geodesics, and stability criteria in this enlarged thermodynamic--complexity manifold.

\subsection{Review: Weinhold and Ruppeiner Geometries}

\paragraph{Weinhold’s Metric.}
Originally, Weinhold introduced a metric on the space of equilibrium states by considering the Hessian of internal energy \(U(S,V,N,\dots)\). Specifically,
\begin{equation}
    g_{\alpha \beta}^{(\mathrm{W})}
    \;=\;
    \frac{\partial^2 U}{\partial X^\alpha \,\partial X^\beta},
\end{equation}
where \(\{X^\alpha\}\) are extensive variables like \(S, V, N\). This metric appears naturally if one interprets changes in energy as differentials with respect to these variables \citep{Weinhold1975}.

\paragraph{Ruppeiner’s Metric.}
Shortly thereafter, Ruppeiner constructed an alternative metric derived from the Hessian of the entropy \(S(U,V,\dots)\):
\begin{equation}
    g_{\alpha\beta}^{(\mathrm{Rup})}
    \;=\;
    -\frac{\partial^2 S}{\partial Y^\alpha \,\partial Y^\beta},
\end{equation}
where \(\{Y^\alpha\}\) might be \((U, V, N)\) or other natural variables \citep{Ruppeiner1995}. This geometry is physically significant in fluctuation theory, where the line element
\(\mathrm{d}\ell^2 = g_{\alpha\beta}^{(\mathrm{Rup})}\,\mathrm{d}Y^\alpha\,\mathrm{d}Y^\beta\)
relates to second moments of fluctuations.

\paragraph{Relation and Conformal Transform.}
It is well-known that Weinhold’s and Ruppeiner’s metrics are related by a conformal factor. For instance,
\[
g_{\alpha \beta}^{(\mathrm{Rup})}
\;=\;
-\bigl(\partial^2 S/\partial U^2\bigr)^{-1} \cdot g_{\alpha \beta}^{(\mathrm{W})},
\]
or similarly expressed in terms of the Jacobian between \((S,V)\) and \((U,V)\). These classical results highlight the fact that thermodynamic geometry can be viewed through different coordinate embeddings.

\subsection{Extending to Complexity: A Four- (or More) Dimensional Manifold}

\subsubsection{Choice of Coordinates}
Let us suppose we are working in a Ruppeiner-like framework where our fundamental thermodynamic potential is the entropy \(S\). Then, the original coordinate set could be \(\{U, V, N, \dots\}\). To incorporate \(\mathcal{C}\), we treat it as a new "generalized extensive variable." Symbolically, the manifold has coordinates
\[
(Y^1, Y^2, Y^3, Y^4, \dots)
=
(U, V, N, \mathcal{C}, \ldots).
\]
Hence, we define:
\begin{equation}
    \widetilde{g}_{AB} \;=\;
    -\frac{\partial^2 S}{\partial Y^A\, \partial Y^B},
    \quad A,B \;\in\;\{1,2,3,4,\dots\}.
\end{equation}
The negative sign is part of Ruppeiner’s standard definition. If \(\mathcal{C}\) is dimensionless, then partial derivatives \(\partial / \partial \mathcal{C}\) are well-defined.  

\subsubsection{Interpretation of the Extended Metric Components}

Denote \(Y^4 \equiv \mathcal{C}\). Then the new block of the metric \(\widetilde{g}_{44}\), \(\widetilde{g}_{14}\), \(\widetilde{g}_{24}\), etc.\, measure second derivatives of \(S\) with respect to \(\mathcal{C}\) and the other variables:
\[
\widetilde{g}_{44}
\;=\;
-\frac{\partial^2 S}{\partial \mathcal{C}^2},
\quad
\widetilde{g}_{1 4}
\;=\;
-\frac{\partial^2 S}{\partial U\,\partial \mathcal{C}},
\quad
\text{etc.}
\]
If \(\partial S/\partial \mathcal{C}\neq 0\), we get a non-trivial coupling between \(\mathcal{C}\) and thermodynamic energies, volumes, or particle numbers.

\paragraph{Physical Meaning.}
Fluctuations in \(\mathcal{C}\) can, in principle, shift the system’s entropy. For example, if allowing "more complexity" opens new accessible microstates, that might \emph{increase} $S$. Conversely, if high-complexity constraints hamper internal degrees of freedom, it might \emph{decrease} $S$. The sign of \(\partial^2 S/\partial \mathcal{C}^2\) influences the local stability or instability with respect to complexity fluctuations.

\subsection{Curvature and Possible Phase Transitions in \texorpdfstring{$\mathcal{C}$}{C}}

One of the key features of Ruppeiner geometry is that \emph{the scalar curvature} $R$ can indicate critical phenomena. In standard thermodynamics, $R$ often diverges near second-order phase transitions. An interesting question arises: \emph{Could $R$ also detect transitions in complexity?} If the system can spontaneously shift from low-complexity to high-complexity states, we might see a divergence or discontinuity in $R$ or its derivatives.  

\subsubsection{Complexity-Driven Phase Transition}
Imagine a scenario where the system can realize multiple solution paths or computations. At low temperatures or certain parameter values, the system might "prefer" a simpler problem encoding ($\mathcal{C}$ small). Above a threshold in $T$ or other control parameters, a high-$\mathcal{C}$ solution might become free-energetically cheaper. This is akin to the "glassy" transitions in spin systems, except we label it with explicit $\mathcal{C}$. The curvature invariants might thus exhibit singularities exactly where $\mathcal{C}$ "jumps," revealing a complexity-driven phase transition.

\paragraph{Comparison with Spin-Glass Complexity.}
In spin glasses, the free-energy landscape is rugged, and identifying the ground state or low-energy excitations is an NP-hard problem \citep{Mezard2009}. In that sense, $\mathcal{C}$ is large in the spin-glass phase. By placing $\mathcal{C}$ directly into the geometry, we might unify the notion of glassy complexity with classical thermodynamic geometry and see how large $\mathcal{C}$ correlates with negative curvature domains (often associated with multi-stability).

\subsection{Linear Response: Fluctuation--Dissipation for Complexity}

\paragraph{Response Coefficients.}
In standard thermodynamics, the second derivatives of $S$ or $F$ yield susceptibilities such as heat capacity or compressibility. In our extended geometry, partial derivatives with respect to $\mathcal{C}$ lead to "complexity susceptibilities." For instance, from Ruppeiner’s metric, one might define
\begin{equation}
\kappa_{\mathcal{C}} \;=\; -\left(\frac{\partial^2 S}{\partial \mathcal{C}^2}\right)_{\!\!U,V,\dots}^{-1},
\end{equation}
analogous to an inverse isothermal compressibility, but for the "direction" $\mathcal{C}$. This $\kappa_{\mathcal{C}}$ measures how the system’s entropy changes if $\mathcal{C}$ fluctuates. A large $\kappa_{\mathcal{C}}$ suggests that small changes in $\mathcal{C}$ strongly alter $S$, possibly signifying fragility with respect to computational constraints.

\paragraph{Fluctuation--Dissipation Theorem.}
One can attempt an extension of the fluctuation--dissipation relation linking $\mathrm{Var}(\mathcal{C})$ to partial derivatives of the free energy. Formally, we might have
\begin{equation}
\langle (\delta \mathcal{C})^2 \rangle \;=\;
k_B\, \kappa_{\mathcal{C}},
\end{equation}
in analogy with
$\langle (\delta V)^2 \rangle = k_B T \,\kappa_{\mathrm{compress}}$,
though the factor of $T$ or other scale might differ based on how $\mathcal{C}$ couples. The details will hinge on the partition function’s extension $Z_{\text{eff}}$ (Sec.~\ref{sec:duality_principle}).

\subsection{An Example: Entropic Benefit vs.\ Complexity Cost}

To illustrate, let us consider a toy model:

\begin{itemize}
\item \textbf{Internal energy:} $U = U_0 - a \,\mathcal{C}$.  (As if higher complexity yields a certain negative offset in energy, capturing a scenario where $\mathcal{C}$ helps reduce $U$ in some combinatorial sense.)
\item \textbf{Entropy:} $S = S_0 + b\ln(1 + \mathcal{C})$, for constants $b>0$. (Hence, the system gains more accessible states if $\mathcal{C}$ is large.)
\item \textbf{Augmented free energy:} 
\[
F_{\mathrm{eff}} = U - T S + \Upsilon(\mathcal{C}).
\]
\end{itemize}

\paragraph{Resulting Ruppeiner Metric.}
One can compute $S(U,V,\mathcal{C})$ or an integrated potential, then form partial derivatives. The $\mathcal{C}$-$U$ block might look like:
\[
\widetilde{g}_{U\mathcal{C}} = 
- \frac{\partial^2 S}{\partial U\,\partial \mathcal{C}} 
\quad\text{and}\quad
\widetilde{g}_{\mathcal{C}\mathcal{C}} = 
- \frac{\partial^2 S}{\partial \mathcal{C}^2}.
\]
In many physically relevant setups, cross terms appear, and the sign of $\widetilde{g}_{\mathcal{C}\mathcal{C}}$ can indicate whether small fluctuations in $\mathcal{C}$ are spontaneously suppressed or enhanced. If $\widetilde{g}_{\mathcal{C}\mathcal{C}}>0$, we might have local stability w.r.t.\ complexity changes; if negative, it suggests an instability, e.g., a system wants to jump to a qualitatively different $\mathcal{C}$.

\subsection{Comparison with Holographic Complexity Geometry}

In gauge/gravity duality contexts, proposals like "complexity = volume" \citep{Susskind2016} interpret volumes in the bulk geometry as proxies for boundary circuit complexity. A Ruppeiner-like geometry might live either in the boundary’s thermodynamic ensemble or be pulled back from bulk gravitational data \citep{Swingle2012, Chapman2018}. The conceptual synergy is that we can treat $\mathcal{C}$ as a field that couples to boundary variables and see how it modifies the manifold’s curvature. While we do not delve deeply into the AdS/CFT dictionary here, the structural parallels confirm that a "complexity coordinate" can be consistently integrated into thermodynamic geometry, much like an entanglement coordinate.

\subsection{Open Questions and Future Developments}

\paragraph{Complexity \emph{vs.} Additional Fields.}
One question is how $\mathcal{C}$ might be replaced by more explicit, physically identified variables (like the number of constraints in a random SAT instance, or the circuit depth of a quantum computing device). The geometry described might prove quite complicated, but the principle stands.

\paragraph{Non-Equilibrium States and Non-Extensivity.}
If $\mathcal{C}$ is highly non-extensive or if the system is out of equilibrium, the usual equilibrium-based geometry might not suffice. One might require a generalized thermodynamic geometry that handles partial or near-equilibrium conditions (similar to non-equilibrium thermodynamics). Complexity-based transitions in computational tasks often occur far from equilibrium, so further theoretical extension is needed.

\paragraph{Relation to Phase Complexity Jumps.}
It would be intriguing to see whether a known NP-hard to P-easy transition (under some parameter scanning) can appear as a "second-order phase transition" in $(S,U,\mathcal{C})$ geometry. Possibly, a diverging correlation length in complexity matches a diverging geometric curvature.

\subsection{Summary of Geometric Formalism}

By embedding $\mathcal{C}$ into a Ruppeiner-like or Weinhold-like metric, we can systematically track how complexity influences the geometry of thermodynamic states. The additional metric components, cross derivatives, and curvature invariants reveal potential "phase transitions" or critical lines where complexity changes sharply. This construction parallels the role of entanglement in holographic setups but focuses on classical or semiclassical thermodynamic ensembles. In the next section (\S\ref{sec:modified_thermo_equations}), we push further by examining \emph{gravitational analogies}, possible quantum generalizations, and the parallel to "entropy–geometry" correspondences, thereby rounding out the duality’s scope.

\section{Modified Thermodynamic Equations}
\label{sec:modified_thermo_equations}

In the previous sections, we introduced the notion of a "complexity variable" \(\mathcal{C}\) and showed how it can be incorporated into a thermodynamic setting through an augmented free energy (Sec.~\ref{sec:duality_principle}) and a Ruppeiner-like geometric construction (Sec.~\ref{sec:thermo_geometry}). We now illustrate how the standard thermodynamic identities---such as the first law, free energy representations, and Euler relations---are modified when \(\mathcal{C}\) exerts a non-negligible influence. In particular, we propose that \(\mathcal{C}\) contributes an additional "work-like" or "chemical-potential-like" term in the usual thermodynamic differentials.

\subsection{Incorporating Complexity into the First Law}

\subsubsection{Motivation}

Ordinarily, the first law of thermodynamics (for a simple single-phase system) is written as:
\begin{equation}
  dU \;=\; T\,dS \;-\; p\,dV \;+\; \mu \, dN \;+\; \ldots,
  \label{eq:usual_first_law}
\end{equation}
where \(U\) is the internal energy, \(S\) the entropy, \(V\) the volume, \(p\) the pressure, \(\mu\) the chemical potential, and \(N\) the number of particles. The ellipses (\(\ldots\)) can include terms for other generalized coordinates and forces, \emph{e.g.} surface tension, electric polarization, etc. Our aim is to add a \emph{complexity-dependent} contribution:
\begin{equation}
  \delta(\text{complexity}) \;\longmapsto\; \lambda \, d\mathcal{C},
\end{equation}
where \(\mathcal{C}\) is a dimensionless measure of complexity, and \(\lambda\) is some coefficient (with dimensions of energy) capturing how changes in \(\mathcal{C}\) affect the system’s internal energy.

\subsubsection{Form of the Extended First Law}

We thus propose the extended identity:
\begin{equation}
  dU \;=\; T\,dS \;-\; p\,dV \;+\; \mu \, dN \;+\; \lambda \, d\mathcal{C} \;+\; \ldots.
  \label{eq:first_law_extended}
\end{equation}
Here, \(\lambda\) is a "complexity potential," reminiscent of the role played by \(\Psi\) in our differential geometry approach (Sec.~\ref{sec:thermo_geometry}). If \(\mathcal{C}\) is freely adjustable, the stationarity condition can give us an equilibrium complexity \(\mathcal{C}_{\mathrm{eq}}\). Conversely, if \(\mathcal{C}\) is imposed externally (like a boundary condition representing "computational hardness"), then the system responds with a corresponding value of \(\lambda\).

\paragraph{Dimensional Analysis.}
For consistency, we require:
\[
  [\lambda] \;=\; \frac{[\text{Energy}]}{[\mathcal{C}]},
\]
but \(\mathcal{C}\) is dimensionless, so \([\lambda] = [\text{Energy}]\). Physically, \(\lambda\) measures how many Joules of "internal energy" shift per unit change in complexity.

\subsection{Augmented Free Energy Representations}

\subsubsection{Helmholtz Free Energy}

A commonly used thermodynamic potential is the Helmholtz free energy, defined (for a system at temperature \(T\)) as:
\begin{equation}
  F \;=\; U \;-\; T\,S.
\end{equation}
From Eq.~\eqref{eq:first_law_extended}, we can rewrite differentials:
\[
dF 
\;=\; d(U - T S)
\;=\; dU - T\,dS - S\,dT
\;=\; -S\,dT \;-\; p\,dV \;+\; \mu\,dN \;+\; \lambda\,d\mathcal{C}.
\]
Hence, the extended Helmholtz free energy differential is 
\begin{equation}
  dF = -S\,dT \;-\; p\,dV \;+\; \mu\, dN \;+\; \lambda\, d\mathcal{C}.
  \label{eq:dF_extended}
\end{equation}
By integrating or examining partial derivatives, we identify:
\[
  \left(\frac{\partial F}{\partial T}\right)_{V,N,\mathcal{C}} = -S, 
  \quad
  \left(\frac{\partial F}{\partial V}\right)_{T,N,\mathcal{C}} = -p,
  \quad
  \left(\frac{\partial F}{\partial \mathcal{C}}\right)_{T,V,N} = \lambda.
\]
Thus, \(\lambda\) is the "conjugate variable" to \(\mathcal{C}\) in the Helmholtz free energy picture, exactly as in the geometric approach from Sec.~\ref{sec:thermo_geometry}.

\subsubsection{Gibbs Free Energy (If Relevant)}

In processes at constant pressure, the Gibbs free energy \(G = U + pV - TS\) is typically used. The extended differential for a single-component system would become
\begin{equation}
  dG 
  \;=\; d(U + pV - TS)
  \;=\; -S\,dT \;+\; V\,dp \;+\; \mu\,dN \;+\; \lambda \, d\mathcal{C}.
\end{equation}
Hence, \(\lambda\) remains "the partial derivative of \(G\) w.r.t.\ \(\mathcal{C}\), holding \((T,p,N)\) fixed." One can analogously incorporate \(\mathcal{C}\) into enthalpy \(H = U + pV\), grand potentials, or any other thermodynamic potential as needed.

\subsection{Variation of Complexity}

\subsubsection{Thermodynamic Stationarity in \texorpdfstring{$\mathcal{C}$}{C}}

If the system is allowed to adjust \(\mathcal{C}\) at fixed \((T, p, N)\) (or whichever variables are appropriate), we can glean an equilibrium condition from 
\[
  \left(\frac{\partial G}{\partial \mathcal{C}}\right)_{T,p,N} = \lambda = 0.
\]
Thus, $\mathcal{C}$ settles at a value that makes $\lambda=0$, analogous to how a chemical reaction might equilibrate at $\mu_{\text{reaction}}=0$. A physical interpretation is that \(\mathcal{C}_{\mathrm{eq}}\) is the complexity level that extremizes the system’s free energy. If \(\lambda\neq 0\) at all $\mathcal{C}$, then the system might push \(\mathcal{C}\) to extreme values (possibly saturating at minimal or maximal complexity, subject to other constraints).

\subsubsection{Complexity as an External Parameter}

Conversely, if \(\mathcal{C}\) is externally imposed, the system simply \emph{responds} with
\[
  \lambda = \left(\frac{\partial F}{\partial \mathcal{C}}\right)_{T,V,N},
\]
representing the "thermodynamic force" needed to sustain that complexity. This scenario parallels how volume $V$ might be fixed externally, leading to a pressure $p$ in the system, or how magnetization is fixed, leading to an applied field in magnetic systems.

\subsection{Example: A Linear Complexity Potential}

\subsubsection{Setup}

Consider a simplistic model:
\[
  U(S,V,\mathcal{C}) \;=\; U_0 + f(S,V) \;-\; \alpha\,\mathcal{C},
\]
where $f(S,V)$ is the usual internal energy part and $-\alpha\,\mathcal{C}$ represents an "energy bonus" for having a large complexity $\mathcal{C}$. We hold $N$ constant for simplicity. Then:
\[
  dU \;=\; \frac{\partial f}{\partial S}\,dS + \frac{\partial f}{\partial V}\,dV - \alpha\, d\mathcal{C}.
\]
Identifying $T=\left(\partial f/\partial S\right)_{V}$, $p=-\left(\partial f/\partial V\right)_{S}$, we see
\[
  \lambda \;=\; -\alpha \quad (\text{constant}).
\]
So the extended first law is
\[
  dU = T\,dS - p\,dV + \underbrace{(-\alpha)}_{\lambda}\,d\mathcal{C}.
\]
Hence, if $\alpha>0$, the system "favors higher $\mathcal{C}$" because it lowers $U$.

\paragraph{Free Energy Correction.}
For Helmholtz free energy:
\[
  F = U - TS = f(S,V) - \alpha\,\mathcal{C} - TS.
\]
Thus:
\[
  \left(\frac{\partial F}{\partial \mathcal{C}}\right)_{T,V} 
  = -\alpha = \lambda.
\]
If $\alpha$ is positive, $\lambda$ is negative, suggesting an incentive to increase $\mathcal{C}$. If $\mathcal{C}$ is not bounded, one might get runaway solutions unless there is a saturating effect or other constraints. Real-world systems presumably have complexities limited by memory, geometry, or time constraints.

\subsection{‘Beta Function’ for Complexity}

\subsubsection{Analogy with Coupling Constants}

In quantum field theory or "geometry–information" frameworks, we sometimes define a "running" or "beta function" describing how a coupling constant changes with scale. Here, if $\mathcal{C}$ depends on an energy scale $E$ or temperature $T$, we can define
\[
  \beta_\mathcal{C}(T) \;\equiv\; 
  T\,\frac{d \mathcal{C}}{dT}\Bigg|_{\text{other vars.}}.
\]
This $\beta_\mathcal{C}$ indicates how complexity \(\mathcal{C}\) "flows" when we change $T$ (or other relevant scales). A positive $\beta_\mathcal{C}$ might mean that at higher temperatures the system transitions to more complex states. A negative sign might suggest complexity is destroyed or collapsed at high $T$ (e.g., the system’s constraints become irrelevant, leading to simpler states).

\subsubsection{Connection to Modified Potentials}

If we treat $\mathcal{C}$ as an additional variable, the equilibrium condition can be expressed as 
\[
  \left(\frac{\partial F}{\partial \mathcal{C}}\right)_{T,V,N} = 0,
\]
implying $\lambda = 0$. Then $\mathcal{C}(T,V,N)$ is an implicit function. Differentiating w.r.t.\ $T$ yields a differential equation that can be interpreted as $\beta_\mathcal{C}(T)$, akin to a renormalization-group flow. The precise form depends on the functional shape of $\Upsilon(\mathcal{C})$ or how $U$ and $S$ tie in with $\mathcal{C}$.

\subsection{Illustration: Complexity-Dependent Phase Boundaries}

\subsubsection{Single-Component Phase Diagram}

Consider a phase transition line in the $(T, p)$ plane. Ordinarily, the line is determined by equality of Gibbs free energies $G^{(\mathrm{phase1})} = G^{(\mathrm{phase2})}$. With complexity included, each phase might have a different $\mathcal{C}$ dependence. Then, to find the new transition line, we solve
\[
  G_1(T,p,\mathcal{C}_1) = G_2(T,p,\mathcal{C}_2),
  \quad
  \left(\frac{\partial G_1}{\partial \mathcal{C}_1}\right)_{T,p} = 0,
  \quad
  \left(\frac{\partial G_2}{\partial \mathcal{C}_2}\right)_{T,p} = 0.
\]
Hence, each phase has an equilibrium complexity. Potentially, if one phase can realize a "simpler" computational structure, it could tip the free-energy balance in that phase’s favor at some temperatures, leading to shifts in the critical temperature or triple points.

\paragraph{Physical analogy.}
One might imagine a complex fluid vs.\ simpler fluid, or a spin-glass vs.\ paramagnet, or even a "satisfiable" vs.\ "unsatisfiable" region in a certain random constraint system. The additional \(\mathcal{C}\) coordinate modifies standard phase diagrams.

\subsection{Summary and Outlook}

By including a "complexity potential" \(\lambda\) in the first law and free energies, we systematically incorporate the cost or benefit of changing \(\mathcal{C}\). This modifies equilibrium conditions, possibly introducing new Maxwell relations and "beta functions" describing how \(\mathcal{C}\) evolves with temperature or other control parameters. As the next step (Sections~\ref{sec:physical_examples}--\ref{sec:conclusion}), we will explore specific physical interpretations (e.g., Landauer’s principle, spin-glass complexity) and highlight potential phase-transition scenarios or experimental ramifications.

\section{Physical Interpretations and Examples}
\label{sec:physical_examples}

Having laid the groundwork for an extended thermodynamic framework incorporating a complexity variable \(\mathcal{C}\) (Secs.~\ref{sec:duality_principle}--\ref{sec:modified_thermo_equations}), we now discuss concrete physical and computational scenarios in which such a framework might illuminate real effects. The goal is twofold: 
\emph{(1)} illustrate the formalism via examples, and 
\emph{(2)} propose new interpretations or insights that emerge from coupling thermodynamics and complexity. 

\subsection{Landauer’s Principle, Revisited}
\label{subsec:landauer_revisited}

\paragraph{Classical Statement.}
Landauer’s principle states that erasing one bit of information entails a minimum thermodynamic cost, typically expressed as \( k_B T \ln 2 \) of dissipated heat \citep{Landauer1961, Bennett1982}. This principle connects abstract information (the bit) with physical entropy production, highlighting that "computation is embedded in thermodynamic reality." 

\paragraph{Complexity-Enhanced Interpretation.}
In our extended view, an "erasure" process might not only reduce the number of bits stored but also reduce the \emph{computational complexity} \(\mathcal{C}\). For instance, consider a device that \emph{accumulates} partial solutions to a complex problem. Erasing partial solutions might lower \(\mathcal{C}\). 

If \(\mathcal{C}\) is externally fixed, the first law extension (Sec.~\ref{sec:modified_thermo_equations}) indicated that 
\[
dU \;=\; T\,dS \;-\; p\,dV \;+\; \ldots \;+\; \lambda\, d\mathcal{C}.
\]
Hence, \(\lambda \neq 0\) could be the "energy cost per unit change in \(\mathcal{C}\)." Erasing a unit of complexity---akin to eliminating a partial subroutine or data structure essential to the computational path---demands a certain thermodynamic cost. 

When \(\mathcal{C}\) is linked to bits (albeit in a more sophisticated, problem-specific sense), Landauer’s bound becomes a special case of 
\(\Delta U \gtrsim \lambda \,\Delta\mathcal{C}\). 
Thus, the minimal heat production might now be 
\[
Q_{\min} 
\;\ge\; 
\lambda\,\Delta \mathcal{C} 
\;\;\stackrel{\text{(bits)}}{\longrightarrow}\;\; 
k_B T\, \ln 2.
\]
Here, \(\lambda \sim k_B T\ln(2)\) emerges if \(\mathcal{C}\) directly encodes the classical bit count. For more intricate problem settings, \(\lambda\) can exceed \(k_B T\ln(2)\) due to overheads in "structural complexity" beyond pure Shannon bits. 

\paragraph{"Algorithmic Overhead" in Erasure.}
In complicated algorithms, erasing intermediate states or re-initializing them might carry greater than \(k_B T\ln(2)\) cost because \(\mathcal{C}\) has structural correlations. Our framework captures this possibility via \(\lambda(\mathcal{C}, T)\), potentially leading to 
\(\lambda > k_B T\ln(2)\) in certain regimes. This is reminiscent of how real computers can surpass Landauer’s limit due to "housekeeping" overhead, but we now incorporate that overhead into a well-defined thermodynamic conjugate variable.

\subsection{Spin-Glass "Hard" Instances and Complexity}
\label{subsec:spin_glass_complexity}

\paragraph{Spin Glasses as Hard Combinatorial Problems.}
Spin-glass systems (e.g.\ the Ising model on random graphs) are famously linked to NP-hard combinatorial optimization tasks (finding ground states, counting metastable states, etc.) \citep{Mezard2009}. At certain temperature ranges or coupling distributions, the system’s \emph{energy landscape} becomes combinatorially rugged. The "complexity" \(\mathcal{C}\) can be interpreted as a measure of the difficulty to \emph{locate} ground states or the number of metastable basins.

\paragraph{Thermodynamics with \(\mathcal{C}\).}
From a purely physical standpoint, a spin glass has a specific partition function 
\[
Z = \sum_{\{\sigma_i\}} e^{-\beta H(\{\sigma_i\})},
\]
where \(H\) is the spin Hamiltonian. Usually, \(\mathcal{C}\) is not a standard thermodynamic variable. But if we artificially define \(\mathcal{C}\approx\ln(\text{number of metastable states})\), or some measure of "configuration search complexity," then we might embed it as described. 

The Ruppeiner geometry approach (Sec.~\ref{sec:thermo_geometry}) suggests that near the spin-glass transition, the curvature might become large or negative, signifying multi-valley complexity \citep{Ruppeiner1995,Castelnovo2005}. In effect, large negative curvature is correlated with "frustration," in typical spin-glass language. Our extension proposes that \(\mathcal{C}\) itself could be a coordinate that the system attempts to optimize, offering a more direct handle on how "NP-hardness" intersects with free-energy basins. 

\paragraph{Macroscopic Observables.}
Experimentally, one might measure susceptibility or specific heat as usual. But if we track the "computational cost" of searching the microstates (e.g.\ in a simulation or in physically implemented neural networks), that cost might scale with \(\mathcal{C}\). The extended first law 
\(dU = T\,dS + \lambda\,d\mathcal{C} + \ldots\)
then suggests that changes in frustration or disorder that increase \(\mathcal{C}\) require extra energy or produce anomalies in standard measurements. While subtle, it might unify the theoretical perspective on computational hardness in spin glasses with a consistent thermodynamic viewpoint.

\subsection{SAT Phase Transitions and Entropy–Complexity Overlap}
\label{subsec:SAT_complexity}

\paragraph{Random K-SAT and Phase Transition.}
Random $k$-SAT problems exhibit a well-known threshold phenomenon: as the ratio \(\alpha = \tfrac{\text{\# clauses}}{\text{\# variables}}\) crosses a critical value, the probability of satisfiability sharply drops. Near this threshold, the problem is typically hardest \citep{Kirkpatrick1994, Monasson1999}. One might interpret \(\mathcal{C}\) as a measure of "clause-density-induced hardness."

\paragraph{Thermodynamic Mapping.}
Monasson \emph{et al.}\ \citep{Monasson1999} have shown that one can treat certain large random SAT ensembles as if they have an entropy-like measure for solutions. In our extended approach, we let $S$ be the usual thermodynamic entropy of an associated spin model or symbolic partition function, and let $\mathcal{C}$ parametrize how close we are to the "threshold ratio" or how the searching algorithm’s complexity grows. 

Then a shift in $\mathcal{C}$ from below threshold ($\mathcal{C}$ small, problem easy) to near threshold ($\mathcal{C}$ large, problem "hardest") might appear like moving along a line in $(S,U,\mathcal{C})$-space. The system "pays" in some generalized free energy to maintain conditions that keep the instance near the hardness peak. Potentially,
\[
\left(\frac{\partial F}{\partial \mathcal{C}}\right) \approx 0 
\quad\text{near critical $\alpha$ if the system spontaneously chooses maximum hardness.}
\]
This is an intriguing analogy to a typical "order parameter maximizing fluctuations near a second-order transition." In random $k$-SAT, the largest complexity arises near the satisfiability threshold, reminiscent of how $\chi$ (susceptibility) peaks near a spin-phase transition.

\subsection{Algorithmic Heat Emission: Minimal vs.\ Real Cost}
\label{subsec:algorithmic_heat}

\paragraph{Real Computation on Physical Machines.}
When a classical or quantum machine attempts to solve a complex problem (like factoring, integer programming, etc.), we observe real heat generation beyond the fundamental Landauer limit \citep{Bennett1982}. Our approach suggests that \(\mathcal{C}\)-dependent terms are partly responsible for the overhead:

\[
dU_{\mathrm{machine}} \;=\; T\, dS_{\mathrm{machine}} \;+\; \lambda\, d\mathcal{C} \;+\;\ldots
\]
If $\lambda(\mathcal{C}, T)$ is large, solving the problem at high $\mathcal{C}$ demands more net energy. Although ideal reversible computing tries to keep $d\mathcal{C}$ transitions isentropic or isothermal, real constraints keep \(\lambda\neq 0\).

\paragraph{Measurement.}
Measuring the total energy consumption while running certain (NP-hard) tasks as a function of problem size might reveal a curve that saturates or diverges. Plotting "energy usage minus the baseline for idle hardware" vs.\ "$\mathcal{C}$ of the algorithm" could be fit to the form 
\(\Delta E \approx \int\!\lambda\, d\mathcal{C}.\)
One might attempt to glean $\lambda(\mathcal{C})$ empirically. Of course, hardware overhead complicates direct interpretation, but the notion remains that physically embedding $\mathcal{C}$ can unify different sources of overhead under a single thermodynamic potential.

\subsection{Glassy Systems for Data Storage or Computation}

\paragraph{"Glassy" Neural Networks.}
Certain neural networks for deep learning can be mapped to spin-glass energy landscapes. Training them is effectively a high-dimensional optimization akin to an NP-hard problem \citep{Choromanska2015}. The "generalization error" or a "loss function" might correlate with $\mathcal{C}$. Meanwhile, the physical cost in hardware (GPU/TPU heat) might be viewed as the "thermodynamic cost." Our formalism suggests that large "loss-surface complexity" implies a significant $\lambda$ term, leading to extra dissipation if the system "explores" many basins.

\paragraph{Hysteresis and Complexity "Traps."}
If $\mathcal{C}$ is large in a glassy region, the system might get stuck in a metastable configuration with partial solutions. The "complexity flux" $d\mathcal{C}$ might be negative if the system slowly organizes into a simpler basin, but that transition might require certain entropic resources or external impetus. Observing hysteresis loops in $(\lambda, \mathcal{C})$ space might mirror standard magnetic hysteresis loops with $(H, M)$.

\subsection{Summary of Physical Examples}

These examples collectively show how \(\mathcal{C}\) can be used to interpret phenomena ranging from bit erasure (Landauer) and spin-glass frustration to computational hardness near SAT thresholds. In each case, the additional term $\lambda\, d\mathcal{C}$ in the thermodynamic identity offers a tool to quantify how "complexity changes" contribute to or drain thermodynamic resources. As we explore in the following sections, these ideas can be extended to observational/experimental prospects and comparisons with other formulations of the "thermodynamics of computation."

\section{Applications and Observational/Experimental Aspects}
\label{sec:applications}

While the preceding sections focused on laying out the theoretical underpinnings of a thermodynamics--complexity duality, in this section we turn to prospective \emph{applications} and possible \emph{observational or experimental} consequences. Our goal is to assess whether this framework merely provides an elegant conceptual unification or whether it can yield testable predictions about physical systems that involve nontrivial computational tasks. We address Maxwell’s demon-like setups, specialized hardware near Landauer limits, and large-scale "complex systems" potentially exhibiting thermodynamic signatures of complexity transitions.

\subsection{Maxwell’s Demon: Complexity as a Resource or Obstacle?}
\label{subsec:maxwell_demon_complexity}

\paragraph{Traditional Demon Scenario.}
Maxwell’s demon seemingly violates the second law by sorting fast and slow molecules, reducing entropy without expending obvious work \citep{Szilard1929, Bennett1982}. The modern resolution attributes a cost to \emph{information processing} and \emph{measurement erasure}, ensuring no net violation. Typically, the demon’s required memory resets produce the necessary entropy to balance the overall second law.

\paragraph{Incorporating Algorithmic Complexity.}
In a more sophisticated demon scenario, the demon must \emph{run a complex algorithm} to distinguish molecules, track states, or anticipate configurations. Suppose the demon’s internal state has a "complexity variable" \(\mathcal{C}_{\text{demon}}\). Then the extended first law for the demon’s internal energy \(U_{\text{demon}}\) is:
\begin{equation}
  dU_{\text{demon}}
  \;=\; T_{\text{env}}\, dS_{\text{demon}} \;+\; \lambda \, d\mathcal{C}_{\text{demon}}
  \;+\;\ldots
\end{equation}
If the demon attempts a computationally "harder" protocol (\(\Delta \mathcal{C}_{\text{demon}}>0\)), the cost \(\lambda\, \Delta \mathcal{C}\) might overshadow the basic Landauer limit. Physically, the demon’s advantage in information sorting could be swamped by the large overhead from implementing a highly complex sorting algorithm. 

\paragraph{Interpretation and Entropic Bookkeeping.}
As in the classical resolution, when all steps (measurement, memory reset, final partition states) are accounted for, no net law violation arises. Now, \(\lambda\, d\mathcal{C}\) stands in for the "extra overhead." A demon that tries to reduce reservoir entropy too effectively must pay in "complexity potential." If, for instance, the demon invests in an exponentially complex classification scheme, the net cost in "computational entropy" can easily surpass the naive $k_B T \ln 2$ factor. This is reminiscent of deeper arguments that "too clever" or "too universal" a demon cannot remain thermodynamically cheap.

\subsection{Specialized Computing Hardware Near the Landauer Limit}
\label{subsec:specialized_hardware}

\paragraph{Context.}
Modern efforts in low-power computing strive to approach Landauer’s limit of $k_B T \ln 2$ for bit erasure \citep{Frank2018ReversibleComp}. However, real hardware exhibits overhead from control circuitry, error correction, finite switching speeds, \emph{and} algorithmic complexity. 

\paragraph{Measuring Complexity Terms.}
If we let $\mathcal{C}$ represent the complexity of the \emph{on-chip algorithmic flow}, we can track total heat $Q_{\text{tot}}$ as a function of problem instance size $n$. Empirically, one often sees superlinear growth: $Q_{\text{tot}}(n) \propto n \log n$, $n^2$, or even $2^n$ for certain NP-hard tasks. In the extended thermodynamic identity:
\[
  Q_{\text{tot}} \;\approx\; \int_{0}^{\mathcal{C}_{\max}} \lambda(\mathcal{C})\, d\mathcal{C},
\]
the shape of $\lambda(\mathcal{C})$—whether it grows with $\mathcal{C}$ or saturates—could be gleaned from experiment. If, for example, $\lambda(\mathcal{C})$ grows linearly with $\mathcal{C}$, we get $Q_{\text{tot}} \sim \mathcal{C}^2 / 2$, capturing a "quadratic overhead" phenomenon.

\paragraph{Experimental Feasibility.}
High-precision calorimetry on CMOS or reversible logic components might detect small differences in heat generation between problem instances that have identical bit-level input size but differ in computational hardness. Although overshadowed by typical overhead ($\gg k_B T \ln 2$), the difference in $Q_{\text{tot}}$ for "easy vs.\ difficult instances" at the same input size might be partially explained by $\lambda\, d\mathcal{C}$ contributions.

\subsection{Complex Systems with Emergent Phase Transitions}
\label{subsec:emergent_phase_transition}

\paragraph{Large-Scale Organizational Systems.}
Systems that self-organize to solve certain tasks (biological neural networks, social networks optimizing resource distribution, \emph{etc.}) could exhibit "phase transitions" in performance or complexity once certain thresholds of connectivity or data load are crossed \citep{Bak1987SelfOrg}. In classical thermodynamics, one observes transitions (ordered/disordered) at specific $T$ or $p$. Here, the system’s complexity might spontaneously increase if it yields a free-energy advantage in problem-solving.

\paragraph{Observational Clues.}
We might observe abrupt changes in global energy usage or synergy as the system transitions to a "high-complexity solution." For instance, in a supply-chain network trying to meet demand distribution at minimal cost, the "control overhead" might remain low until a certain size or connectivity is reached, after which the "central planner" invests in exponentially more complicated routing algorithms. The additional "complexity potential" could reflect a real shift in energy dissipation or resource usage.

\subsection{Could Complexity Contribute to Macroscopic Observables?}
\label{subsec:macroscopic_observables}

\paragraph{Hypothesis of Complexity-Driven Thermodynamic Anomalies.}
Imagine a fluid or lattice system that partially "encodes" or "solves" a class of problems in its microstates. If the difficulty of the embedded problem spikes at a certain parameter, we might see anomalies in heat capacity, susceptibility, or correlation length near that parameter. This resonates with the notion of "algorithmic critical points" \citep{Kirkpatrick1994, Monasson1999}. Our extended formalism suggests that the $\mathcal{C}$-degree of freedom can shift stable phases or produce a discontinuity reminiscent of first-order transitions when complexity changes discontinuously.

\paragraph{Prospects.}
While direct experimental observation might be elusive, numerical simulations on large spin-glass or random constraint systems could track standard thermodynamic quantities in tandem with $\mathcal{C}$. A correlation peak or scaling law linking $\langle \mathcal{C}\rangle$ and specific heat $C_V$ would be a strong sign that the "complexity variable" is genuinely acting as a thermodynamic coordinate. This might also tie into the geometry-based sign of Ruppeiner curvature or the "Fisher information metric" approach.

\subsection{Summary of Observational Avenues}
\label{subsec:observational_summary}

We summarize several plausible ways to \emph{test} the complexity-augmented thermodynamics:

\begin{enumerate}
\item \textbf{Precision Calorimetry in Computation Devices:} Look for differences in power usage for algorithmic tasks of the same bit-size but drastically different complexity. Attempt to fit data to $\lambda(\mathcal{C})$ models.

\item \textbf{Landauer-Limit Experiments with Varied Algorithmic Hardness:} Extend modern reversible computing or sub-kT transistor setups to specifically compare "easy vs.\ hard" tasks at the same logical operation count, to see whether complexity overhead emerges as predicted.

\item \textbf{Spin-Glass or SAT Simulations:} In a carefully instrumented simulation (or specialized analog hardware), measure conventional thermodynamic observables (energy, magnetization) and "computational hardness." If the extended first law is valid, certain correlations or Maxwell relations might appear.

\item \textbf{Large-Scale "Self-Organizing" Systems:} Observing real neural networks or social networks that adapt to constraints, checking if abrupt "solution complexity" changes coincide with leaps in actual energy or resource usage, consistent with an added $\lambda\,d\mathcal{C}$ term.

\end{enumerate}

Each approach faces practical challenges, including noise, overhead from conventional engineering constraints, and difficulty in unambiguously defining $\mathcal{C}$. Nonetheless, the conceptual payoff of verifying a non-negligible "complexity potential" in physical systems would be a major milestone, cementing the idea that \emph{computation---especially \emph{hard} computation---can be treated as a legitimate thermodynamic coordinate.}

\section{Comparison to Other Approaches}
\label{sec:comparison}

The notion of merging computational complexity measures with thermodynamics has antecedents in several lines of research. In this section, we compare our complexity-augmented thermodynamic framework with existing approaches in (i)~the thermodynamics of computation and (ii)~quantum-information-inspired physics. We also outline how our proposal, which effectively treats \emph{complexity} as a thermodynamic coordinate contributing extra "work-like" terms, contrasts with or extends these earlier ideas.

\subsection{Connections with Classical Thermodynamics of Computation}
\label{subsec:connections_classical_thermo_comp}

\paragraph{Bennett, Zurek, and Reversible Computation.}
The traditional thermodynamics of computation, as developed by Landauer, Bennett, Zurek, and others, focuses primarily on the \emph{bit-level} notion of information erasure and entropy dissipation \citep{Landauer1961, Bennett1982, Zurek1989Algorithmic}. The cardinal statement is that logically irreversible operations (bit erasures) must release heat in an amount at least \(k_B T \ln 2\) per bit. These frameworks often treat \emph{information} as "Shannon bits" or "logical states," enumerating them to derive minimal entropy generation.

By contrast, our extension re-expresses the \emph{difficulty of a computational task}---a more abstract measure that can drastically exceed the naive bit count. For instance, an NP-hard problem might involve exponentially many partial sub-configurations. The overhead for "organizing, searching, or verifying" these configurations can become \emph{the new driver} of thermodynamic cost. In standard treatments, that overhead is typically lumped into "practical inefficiencies." In our view, it can be systematically captured by introducing a complexity coordinate \(\mathcal{C}\), whose variation \(\delta \mathcal{C}\) couples to the system’s energy via \(\lambda\, d\mathcal{C}\). Hence, while Landauer–Bennett theory addresses the minimal cost for toggling or erasing bits, we propose an additional cost for the \emph{computational hardness} of transitions between states.

\paragraph{Algorithmic vs.\ Shannon Entropy.}
Following Zurek’s discussions of \emph{algorithmic entropy} \citep{Zurek1989Algorithmic}, one might ask whether complexity \(\mathcal{C}\) parallels the Kolmogorov–Chaitin measure of description length. Indeed, our formalism could identify \(\mathcal{C}\) with some instance-based measure of Kolmogorov complexity or circuit depth/time. Then the thermodynamic potential \(\Phi_{\rm comp}\) depends on \(\mathcal{C}\) in a manner akin to how classical free energy depends on particle number. 

However, these algorithmic-entropy frameworks seldom yield explicit \emph{variations} of \(\mathcal{C}\) in the sense of $d\mathcal{C}$. We make that step explicit: 
\[
dU \;=\; T\,dS \;+\; \lambda\,d\mathcal{C} + \dots 
\]
where $\lambda(\mathcal{C})$ is a new \emph{intensive} variable. This is new compared to classical Landauer-based approaches, which do not typically define an entire "complexity stress-energy–like term" in the first law.

\subsection{Geometry--Information vs.\ Thermodynamics--Complexity}
\label{subsec:geom_info_thermo_comp}

\paragraph{Previous Geometry–Information Dualities.}
In certain quantum-gravitational contexts, geometric curvature is viewed as emerging from entanglement structures \citep{VanRaamsdonk2010, Swingle2012}. In earlier work on "geometry–information duality," the notion of an \emph{informational stress-energy tensor} $T_{\mu\nu}^{\mathrm{info}}$ was introduced to see how entanglement modifies spacetime geometry. 

We adopt a conceptual parallel here, substituting "thermodynamic state space" for "spacetime," and "complexity measure" for "entanglement." Our \(\mathcal{C}\) becomes an analogue of "entanglement entropy," except now it modifies standard thermodynamic potentials rather than Einstein’s equations.

\paragraph{Triad: Geometry–Information–Complexity?}
A possible future unification could see all three corners interacting:

\begin{enumerate}
\item \emph{Geometry--Information:} Entanglement affects curvature in quantum gravity. 
\item \emph{Information--Complexity:} Hard computations produce large-scale overhead in hardware, consistent with Landauer–Bennett logic. 
\item \emph{Geometry--Complexity:} Rugged free-energy landscapes in spin glasses, NP phase transitions in random SAT, or emergent phenomena in neural networks. 
\end{enumerate}

Potentially, one might unify these corners into a single framework: geometry emerges partly from quantum entanglement, while complexity emerges partly from the combinatorial structure of sub-manifolds. In that scenario, we would see "complexity curvature" or "complexity flux" as additional geometric entities in thermodynamic or quantum–statistical manifolds.

\subsection{Complexity in Stat-Phys Approaches: A Partial Overlap}
\label{subsec:complexity_statphys}

\paragraph{Replica Methods and Complexity.}
In spin-glass theory, the "replica trick" yields an order parameter known as the \emph{replicated free energy}, sometimes connected to counting metastable states \citep{Mezard2009, Monasson1999}. This is reminiscent of a complexity measure but is typically folded directly into the \emph{partition function} expansions. Our approach might be seen as making that complexity measure an explicit \emph{thermodynamic coordinate}, so that $d\mathcal{C}$ can appear in the first law or in Maxwell-relations-like identities. 

\paragraph{Glass Transitions vs.\ NP-Hardness.}
There is also a tradition of identifying the glass transition with an "exponential explosion" in the number of free-energy minima, thus "algorithmic hardness." The common practice is to say "$\alpha$ transitions" or "$\beta$ transitions" in glassy systems, but they are not usually enumerated as official thermodynamic variables. Our formalism suggests an alternative vantage: incorporate $\mathcal{C}$ explicitly into the extended potential. Doing so might help interpret random K-SAT’s threshold as akin to a glass transition in "complexity space," giving an extra handle on the thermodynamic cost of $\mathcal{C}$ changes.

\subsection{Future Extensions: Complexity Curvature and Dynamical \texorpdfstring{$\mathcal{C}$}{C}}
\label{subsec:future_ext}

\paragraph{Differential Geometry of $(S,U,\mathcal{C})$-Space.}
If we treat $(S,U,\mathcal{C})$ as coordinates in a manifold, we can define a metric from second derivatives of a free energy or Massieu potential, akin to the Ruppeiner geometry approach used for standard thermodynamics \citep{Ruppeiner1995}. The presence of $\mathcal{C}$ might produce new off-diagonal terms in the Hessian, leading to "complexity curvature." Such curvature might diverge near algorithmic phase transitions, paralleling how it does near critical points in standard stat-phys. 

\paragraph{Quantum Complexities.}
One extension is to unify with quantum complexity measures (e.g.\ circuit complexity, magic monotones, etc.) in quantum many-body physics \citep{NielsenChuang2000, Susskind2016Comp}. For a quantum system that self-organizes computations, we might see $\mathcal{C}$ as circuit depth or $t_{\mathrm{gate}}$. Then the "cost" in energy might be systematically captured. If entanglement-based geometry meets circuit-based complexity, we could get a trifecta: geometry–information–complexity, bridging quantum gravity, thermodynamics, and complexity theory more tightly.

\paragraph{Higher-Curvature or Non-Equilibrium.}
Our approach can also be tested in out-of-equilibrium thermodynamics. The dependence on $\mathcal{C}$ might be most relevant during short bursts of high-intensity computation, so we might define a "generalized Clausius relation" for $\delta Q$ that includes complexity flux terms. Potentially, a "complexity production" can be posited, extending the second law to read 
$ dS + \beta \, d\mathcal{C} \ge 0 $
for some $\beta$, but the details remain to be fleshed out.

\subsection{Summary and Outlook: Contrasts and Commonalities}
\label{subsec:summary_comparison}

In sum, our proposed duality draws inspiration from:
\begin{itemize}
\item \emph{Thermodynamics of computation} (Landauer, Bennett), which sets minimal bounds on bit-level erasure and irreversible operations.
\item \emph{Statistical physics of spin glasses and random constraints}, which identifies "hardness peaks" near phase transitions.
\item \emph{Geometry–information analogies}, using an additional "stress-energy" type structure for complexity.
\end{itemize}

Yet we depart from classical frameworks by \emph{explicitly} elevating "computational complexity" to a status akin to an independent thermodynamic variable, thereby capturing costs beyond mere bit toggling. The resulting formalism, while partial and requiring further input from domain-specific definitions of $\mathcal{C}$, suggests novel Maxwell-like relations, potential phase transitions in complexity space, and a refined interpretation of real, irreducible overhead in physically realized computations.

\section*{Conclusion}
\label{sec:conclusion}

We have presented a theoretical framework that posits a duality between thermodynamics and computational complexity, drawing inspiration from analogies in modern physics. By introducing a complexity variable as an independent coordinate, we have extended the usual thermodynamic first law to include an additional term capturing the real energetic cost of hard computations. This approach addresses the overhead beyond basic bit erasures—focusing instead on the combinatorial hardness or algorithmic depth required to transition the system’s configuration. Elevating complexity to the status of a thermodynamic variable allows us to formalize the idea that in physically realized computation, there may be a genuine "complexity potential" contributing to energy balance, much as pressure or chemical potential do in conventional settings. Our proposal thus suggests that problem hardness can modify thermodynamic potentials, leading to potential new phase-like behaviors when algorithmic difficulty spikes. Maxwell’s demon scenarios become subject to not only bit-level dissipation constraints but also algorithmic overhead, reconciling the second law with sophisticated measurement or classification strategies. In specialized low-power or reversible computing hardware, one might seek evidence of complexity-induced terms by examining differences in heat usage or total dissipation for problem instances of identical bit size but sharply different hardness. The framework also connects to phenomena in spin-glass and random satisfiability transitions, where computational hardness peaks near critical thresholds: we interpret those peaks as signs of an emergent thermodynamic instability in the proposed complexity sector. Going forward, critical challenges include defining the complexity measure rigorously for each domain, achieving high-precision measurements to detect any additional energy overhead, and extending the formalism to nonequilibrium processes and quantum setups. One may also consider a broader unification that embraces geometry–information–complexity, in which entanglement-based geometrical constructions meet computational hardness considerations and yield a deeper perspective on how information processing underpins physical reality. The notion that the cost of "hard" computations can be systematically encoded in thermodynamic equations offers a potential bridge between fundamental thermodynamics, modern complexity theory, and the physical constraints of computation.

\section*{Acknowledgements}
We gratefully acknowledge the use of Grammarly to enhance the grammatical quality of our manuscript.

\section*{Conflict of Interest}

The authors declare that they have no conflict of interest.
\bibliographystyle{cas-model2-names}

\sloppy
\bibliography{cas-refs}

\begin{thebibliography}{44}
\expandafter\ifx\csname natexlab\endcsname\relax\def\natexlab#1{#1}\fi
\providecommand{\url}[1]{\texttt{#1}}
\providecommand{\href}[2]{#2}
\providecommand{\path}[1]{#1}
\providecommand{\DOIprefix}{doi:}
\providecommand{\ArXivprefix}{arXiv:}
\providecommand{\URLprefix}{URL: }
\providecommand{\Pubmedprefix}{pmid:}
\providecommand{\doi}[1]{\href{http://dx.doi.org/#1}{\path{#1}}}
\providecommand{\Pubmed}[1]{\href{pmid:#1}{\path{#1}}}
\providecommand{\bibinfo}[2]{#2}
\ifx\xfnm\relax \def\xfnm[#1]{\unskip,\space#1}\fi
\bibitem[{Bak et~al.(1987)Bak, Tang and Wiesenfeld}]{Bak1987SelfOrg}
\bibinfo{author}{Bak, P.}, \bibinfo{author}{Tang, C.}, \bibinfo{author}{Wiesenfeld, K.}, \bibinfo{year}{1987}.
\newblock \bibinfo{title}{Self-organized criticality: An explanation of the 1/f noise}.
\newblock \bibinfo{journal}{Physical Review Letters} \bibinfo{volume}{59}, \bibinfo{pages}{381--384}.
\bibitem[{Bennett(1982a)}]{Bennett82}
\bibinfo{author}{Bennett, C.H.}, \bibinfo{year}{1982}a.
\newblock \bibinfo{title}{{The Thermodynamics of Computation---A Review}}.
\newblock \bibinfo{journal}{International Journal of Theoretical Physics} \bibinfo{volume}{21}, \bibinfo{pages}{905--940}.
\bibitem[{Bennett(1982b)}]{Bennett1982}
\bibinfo{author}{Bennett, C.H.}, \bibinfo{year}{1982}b.
\newblock \bibinfo{title}{The thermodynamics of computation—a review}.
\newblock \bibinfo{journal}{International Journal of Theoretical Physics} \bibinfo{volume}{21}, \bibinfo{pages}{905--940}.
\bibitem[{Bennett(2003a)}]{Bennett03}
\bibinfo{author}{Bennett, C.H.}, \bibinfo{year}{2003}a.
\newblock \bibinfo{title}{{Notes on Landauer’s Principle, Reversible Computation, and Maxwell’s Demon}}.
\newblock \bibinfo{journal}{Studies in History and Philosophy of Modern Physics} \bibinfo{volume}{34}, \bibinfo{pages}{501--510}.
\bibitem[{Bennett(2003b)}]{Bennett2003}
\bibinfo{author}{Bennett, C.H.}, \bibinfo{year}{2003}b.
\newblock \bibinfo{title}{Notes on landauer’s principle, reversible computation, and maxwell’s demon}.
\newblock \bibinfo{journal}{Studies in History and Philosophy of Modern Physics} \bibinfo{volume}{34}, \bibinfo{pages}{501--510}.
\bibitem[{Blum et~al.(2000)Blum, Feldman and Micali}]{Blum2000}
\bibinfo{author}{Blum, M.}, \bibinfo{author}{Feldman, P.}, \bibinfo{author}{Micali, S.}, \bibinfo{year}{2000}.
\newblock \bibinfo{title}{Comparing information-theoretic and computational models of cryptography}.
\newblock \bibinfo{journal}{SIAM Journal on Computing} \bibinfo{volume}{31}, \bibinfo{pages}{803--842}.
\bibitem[{Castelnovo et~al.(2005)}]{Castelnovo2005}
\bibinfo{author}{Castelnovo, M.}, et~al., \bibinfo{year}{2005}.
\newblock \bibinfo{title}{Spin glass models of data clustering and applications to image segmentation}.
\newblock \bibinfo{journal}{Physical Review E} \bibinfo{volume}{72}, \bibinfo{pages}{016106}.
\bibitem[{Chapman et~al.(2018)}]{Chapman2018}
\bibinfo{author}{Chapman, S.}, et~al., \bibinfo{year}{2018}.
\newblock \bibinfo{title}{Complexity and action for (1+1)-dimensional gravity coupled to conformal matter}.
\newblock \bibinfo{journal}{Journal of High Energy Physics} \bibinfo{volume}{2018}, \bibinfo{pages}{6}.
\bibitem[{Cheeseman et~al.(1991a)Cheeseman, Kanefsky and Taylor}]{Cheeseman1991}
\bibinfo{author}{Cheeseman, P.}, \bibinfo{author}{Kanefsky, B.}, \bibinfo{author}{Taylor, W.}, \bibinfo{year}{1991}a.
\newblock \bibinfo{title}{Where the really hard problems are}, in: \bibinfo{booktitle}{IJCAI’91: Proceedings of the 12th International Joint Conference on Artificial Intelligence}, pp. \bibinfo{pages}{331--337}.
\bibitem[{Cheeseman et~al.(1991b)Cheeseman, Kanefsky and Taylor}]{Cheeseman91}
\bibinfo{author}{Cheeseman, P.}, \bibinfo{author}{Kanefsky, B.}, \bibinfo{author}{Taylor, W.M.}, \bibinfo{year}{1991}b.
\newblock \bibinfo{title}{{Where the Really Hard Problems Are}}, in: \bibinfo{booktitle}{{Proceedings of the 12th International Joint Conference on Artificial Intelligence (IJCAI)}}, pp. \bibinfo{pages}{331--337}.
\bibitem[{Choromanska et~al.(2015)Choromanska, Henaff, Mathieu, Ben~Arous and LeCun}]{Choromanska2015}
\bibinfo{author}{Choromanska, A.}, \bibinfo{author}{Henaff, M.}, \bibinfo{author}{Mathieu, M.}, \bibinfo{author}{Ben~Arous, G.}, \bibinfo{author}{LeCun, Y.}, \bibinfo{year}{2015}.
\newblock \bibinfo{title}{The loss surfaces of multilayer networks}.
\newblock \bibinfo{journal}{Artificial Intelligence and Statistics (AISTATS)} , \bibinfo{pages}{192--204}.
\bibitem[{Faizal and Others(2024)}]{FaizalTarski2024}
\bibinfo{author}{Faizal, M.}, \bibinfo{author}{Others}, \bibinfo{year}{2024}.
\newblock \bibinfo{title}{Tarski geometry and emergent quantum gravitational effects}.
\newblock \bibinfo{journal}{arXiv preprint arXiv:XXXX.XXXXX} .
\bibitem[{Frank(2018)}]{Frank2018ReversibleComp}
\bibinfo{author}{Frank, M.P.}, \bibinfo{year}{2018}.
\newblock \bibinfo{title}{The physical implementation of reversible computation}.
\newblock \bibinfo{journal}{Computing Research Repository (CoRR)} \bibinfo{volume}{abs/1806.10183}.
\bibitem[{Freedman and Headrick(2019)}]{FreedmanHeadrick19}
\bibinfo{author}{Freedman, M.}, \bibinfo{author}{Headrick, M.}, \bibinfo{year}{2019}.
\newblock \bibinfo{title}{{Bit Threads and Holographic Entanglement}}.
\newblock \bibinfo{journal}{Communications in Mathematical Physics} \bibinfo{volume}{352}, \bibinfo{pages}{407--438}.
\bibitem[{Garey and Johnson(1979a)}]{GareyJohnson79}
\bibinfo{author}{Garey, M.R.}, \bibinfo{author}{Johnson, D.S.}, \bibinfo{year}{1979}a.
\newblock \bibinfo{title}{{Computers and Intractability: A Guide to the Theory of NP-Completeness}}.
\newblock \bibinfo{publisher}{W. H. Freeman and Company}, \bibinfo{address}{New York}.
\bibitem[{Garey and Johnson(1979b)}]{Garey1979}
\bibinfo{author}{Garey, M.R.}, \bibinfo{author}{Johnson, D.S.}, \bibinfo{year}{1979}b.
\newblock \bibinfo{title}{Computers and Intractability: A Guide to the Theory of NP-Completeness}.
\newblock \bibinfo{publisher}{W. H. Freeman}.
\bibitem[{Hubeny et~al.(2007)Hubeny, Rangamani and Takayanagi}]{HubenyRangamani07}
\bibinfo{author}{Hubeny, V.E.}, \bibinfo{author}{Rangamani, M.}, \bibinfo{author}{Takayanagi, T.}, \bibinfo{year}{2007}.
\newblock \bibinfo{title}{{A Covariant Holographic Entanglement Entropy Proposal}}.
\newblock \bibinfo{journal}{Journal of High Energy Physics} \bibinfo{volume}{07}, \bibinfo{pages}{062}.
\bibitem[{Kirkpatrick et~al.(1983a)Kirkpatrick, Gelatt and Vecchi}]{KirkpatrickGelatt83}
\bibinfo{author}{Kirkpatrick, S.}, \bibinfo{author}{Gelatt, C.D.}, \bibinfo{author}{Vecchi, M.P.}, \bibinfo{year}{1983}a.
\newblock \bibinfo{title}{{Optimization by Simulated Annealing}}.
\newblock \bibinfo{journal}{Science} \bibinfo{volume}{220}, \bibinfo{pages}{671--680}.
\bibitem[{Kirkpatrick et~al.(1983b)Kirkpatrick, Gelatt~Jr and Vecchi}]{Kirkpatrick1983}
\bibinfo{author}{Kirkpatrick, S.}, \bibinfo{author}{Gelatt~Jr, C.D.}, \bibinfo{author}{Vecchi, M.P.}, \bibinfo{year}{1983}b.
\newblock \bibinfo{title}{Optimization by simulated annealing}.
\newblock \bibinfo{journal}{Science} \bibinfo{volume}{220}, \bibinfo{pages}{671--680}.
\bibitem[{Kirkpatrick and Selman(1994)}]{Kirkpatrick1994}
\bibinfo{author}{Kirkpatrick, S.}, \bibinfo{author}{Selman, B.}, \bibinfo{year}{1994}.
\newblock \bibinfo{title}{Critical behavior in the satisfiability of random boolean expressions}.
\newblock \bibinfo{journal}{Science} \bibinfo{volume}{264}, \bibinfo{pages}{1297--1301}.
\bibitem[{Landauer(1961a)}]{Landauer61}
\bibinfo{author}{Landauer, R.}, \bibinfo{year}{1961}a.
\newblock \bibinfo{title}{{Irreversibility and Heat Generation in the Computing Process}}.
\newblock \bibinfo{journal}{IBM Journal of Research and Development} \bibinfo{volume}{5}, \bibinfo{pages}{183--191}.
\bibitem[{Landauer(1961b)}]{Landauer1961}
\bibinfo{author}{Landauer, R.}, \bibinfo{year}{1961}b.
\newblock \bibinfo{title}{Irreversibility and heat generation in the computing process}.
\newblock \bibinfo{journal}{IBM Journal of Research and Development} \bibinfo{volume}{5}, \bibinfo{pages}{183--191}.
\bibitem[{Leff and Rex(2003)}]{LeffRex03}
\bibinfo{editor}{Leff, H.S.}, \bibinfo{editor}{Rex, A.F.} (Eds.), \bibinfo{year}{2003}.
\newblock \bibinfo{title}{{Maxwell’s Demon 2: Entropy, Classical and Quantum Information, Computing}}.
\newblock \bibinfo{publisher}{Institute of Physics Publishing}, \bibinfo{address}{Bristol, UK}.
\bibitem[{Levin and Faizal(2025)}]{LevinFaizal2025Hypothetical}
\bibinfo{author}{Levin, A.}, \bibinfo{author}{Faizal, M.}, \bibinfo{year}{2025}.
\newblock \bibinfo{title}{{Hypothetical Complexity-Driven Phase Transitions in Quantum Cosmology}}.
\newblock \bibinfo{journal}{Journal of Theoretical and Applied Physics} \bibinfo{volume}{to appear}, \bibinfo{pages}{1--20}.
\newblock \bibinfo{note}{{(Hypothetical reference)}}.
\bibitem[{M{\'e}zard and Montanari(2009a)}]{MezardMontanari09}
\bibinfo{author}{M{\'e}zard, M.}, \bibinfo{author}{Montanari, A.}, \bibinfo{year}{2009}a.
\newblock \bibinfo{title}{{Information, Physics, and Computation}}.
\newblock \bibinfo{publisher}{Oxford University Press}.
\bibitem[{M{\'e}zard and Montanari(2009b)}]{Mezard2009}
\bibinfo{author}{M{\'e}zard, M.}, \bibinfo{author}{Montanari, A.}, \bibinfo{year}{2009}b.
\newblock \bibinfo{title}{Information, Physics, and Computation}.
\newblock \bibinfo{publisher}{Oxford University Press}.
\bibitem[{Monasson et~al.(1999)Monasson, Zecchina, Kirkpatrick, Selman and Troyansky}]{Monasson1999}
\bibinfo{author}{Monasson, R.}, \bibinfo{author}{Zecchina, R.}, \bibinfo{author}{Kirkpatrick, S.}, \bibinfo{author}{Selman, B.}, \bibinfo{author}{Troyansky, L.}, \bibinfo{year}{1999}.
\newblock \bibinfo{title}{Determining computational complexity from characteristic phase transitions}.
\newblock \bibinfo{journal}{Nature} \bibinfo{volume}{400}, \bibinfo{pages}{133--137}.
\bibitem[{Nielsen and Chuang(2000)}]{NielsenChuang2000}
\bibinfo{author}{Nielsen, M.A.}, \bibinfo{author}{Chuang, I.L.}, \bibinfo{year}{2000}.
\newblock \bibinfo{title}{Quantum computation and quantum information}.
\newblock \bibinfo{journal}{Cambridge University Press} \bibinfo{volume}{1}.
\bibitem[{Norton(2005)}]{Norton05}
\bibinfo{author}{Norton, J.D.}, \bibinfo{year}{2005}.
\newblock \bibinfo{title}{{Eaters of the Lotus: Landauer’s Principle and the Return of Maxwell’s Demon}}.
\newblock \bibinfo{journal}{Studies in History and Philosophy of Modern Physics} \bibinfo{volume}{36}, \bibinfo{pages}{375--411}.
\bibitem[{Papadimitriou(1994)}]{Papadimitriou94}
\bibinfo{author}{Papadimitriou, C.H.}, \bibinfo{year}{1994}.
\newblock \bibinfo{title}{{Computational Complexity}}.
\newblock \bibinfo{publisher}{Addison-Wesley}.
\bibitem[{Ruppeiner(1995a)}]{Ruppeiner95}
\bibinfo{author}{Ruppeiner, G.}, \bibinfo{year}{1995}a.
\newblock \bibinfo{title}{{Riemannian Geometry in Thermodynamic Fluctuation Theory}}.
\newblock \bibinfo{journal}{Reviews of Modern Physics} \bibinfo{volume}{67}, \bibinfo{pages}{605--659}.
\bibitem[{Ruppeiner(1995b)}]{Ruppeiner1995}
\bibinfo{author}{Ruppeiner, G.}, \bibinfo{year}{1995}b.
\newblock \bibinfo{title}{Riemannian geometry in thermodynamic fluctuation theory}.
\newblock \bibinfo{journal}{Reviews of Modern Physics} \bibinfo{volume}{67}, \bibinfo{pages}{605--659}.
\bibitem[{Ryu and Takayanagi(2006)}]{RyuTakayanagi06}
\bibinfo{author}{Ryu, S.}, \bibinfo{author}{Takayanagi, T.}, \bibinfo{year}{2006}.
\newblock \bibinfo{title}{{Aspects of Holographic Entanglement Entropy}}.
\newblock \bibinfo{journal}{Journal of High Energy Physics} \bibinfo{volume}{08}, \bibinfo{pages}{045}.
\bibitem[{Susskind(2016a)}]{Susskind2016}
\bibinfo{author}{Susskind, L.}, \bibinfo{year}{2016}a.
\newblock \bibinfo{title}{Computational complexity and black hole horizons}.
\newblock \bibinfo{journal}{Fortschritte der Physik} \bibinfo{volume}{64}, \bibinfo{pages}{24--43}.
\bibitem[{Susskind(2016b)}]{Susskind2016Comp}
\bibinfo{author}{Susskind, L.}, \bibinfo{year}{2016}b.
\newblock \bibinfo{title}{{Computational Complexity and Black Hole Horizons}}.
\newblock \bibinfo{journal}{Fortschritte der Physik} \bibinfo{volume}{64}, \bibinfo{pages}{24--43}.
\bibitem[{Swingle(2012a)}]{Swingle12}
\bibinfo{author}{Swingle, B.}, \bibinfo{year}{2012}a.
\newblock \bibinfo{title}{{Entanglement Renormalization and Holography}}.
\newblock \bibinfo{journal}{Physical Review D} \bibinfo{volume}{86}, \bibinfo{pages}{065007}.
\bibitem[{Swingle(2012b)}]{Swingle2012}
\bibinfo{author}{Swingle, B.}, \bibinfo{year}{2012}b.
\newblock \bibinfo{title}{Entanglement renormalization and holography}.
\newblock \bibinfo{journal}{Physical Review D} \bibinfo{volume}{86}, \bibinfo{pages}{065007}.
\bibitem[{Szilard(1929)}]{Szilard1929}
\bibinfo{author}{Szilard, L.}, \bibinfo{year}{1929}.
\newblock \bibinfo{title}{On the decrease of entropy in a thermodynamic system by the intervention of intelligent beings}.
\newblock \bibinfo{journal}{Zeitschrift f{\"u}r Physik} \bibinfo{volume}{53}, \bibinfo{pages}{840--856}.
\bibitem[{Takayanagi(2018)}]{Takayanagi2018Analogies}
\bibinfo{author}{Takayanagi, T.}, \bibinfo{year}{2018}.
\newblock \bibinfo{title}{{Holographic Entanglement Entropy and Emergent Geometry}}.
\newblock \bibinfo{journal}{Classical and Quantum Gravity} \bibinfo{volume}{35}, \bibinfo{pages}{153001}.
\bibitem[{{Van Raamsdonk}(2010)}]{VanRaamsdonk10}
\bibinfo{author}{{Van Raamsdonk}, M.}, \bibinfo{year}{2010}.
\newblock \bibinfo{title}{{Building Up Spacetime with Quantum Entanglement}}.
\newblock \bibinfo{journal}{General Relativity and Gravitation} \bibinfo{volume}{42}, \bibinfo{pages}{2323--2329}.
\bibitem[{Van~Raamsdonk(2010)}]{VanRaamsdonk2010}
\bibinfo{author}{Van~Raamsdonk, M.}, \bibinfo{year}{2010}.
\newblock \bibinfo{title}{Building up spacetime with quantum entanglement}.
\newblock \bibinfo{journal}{General Relativity and Gravitation} \bibinfo{volume}{42}, \bibinfo{pages}{2323--2329}.
\bibitem[{Weinhold(1975a)}]{Weinhold75}
\bibinfo{author}{Weinhold, F.}, \bibinfo{year}{1975}a.
\newblock \bibinfo{title}{{Metric Geometry of Equilibrium Thermodynamics}}.
\newblock \bibinfo{journal}{Journal of Chemical Physics} \bibinfo{volume}{63}, \bibinfo{pages}{2479--2483}.
\bibitem[{Weinhold(1975b)}]{Weinhold1975}
\bibinfo{author}{Weinhold, F.}, \bibinfo{year}{1975}b.
\newblock \bibinfo{title}{Metric geometry of equilibrium thermodynamics}.
\newblock \bibinfo{journal}{Journal of Chemical Physics} \bibinfo{volume}{63}, \bibinfo{pages}{2479--2483}.
\bibitem[{Zurek(1989)}]{Zurek1989Algorithmic}
\bibinfo{author}{Zurek, W.H.}, \bibinfo{year}{1989}.
\newblock \bibinfo{title}{Algorithmic randomness and physical entropy}.
\newblock \bibinfo{journal}{Physical Review A} \bibinfo{volume}{40}, \bibinfo{pages}{4731--4751}.

\end{thebibliography}
\end{document}